\documentstyle[12pt,epsf]{article}
\epsfverbosetrue
\def\figlabel#1{\xdef#1{\thefigure}}
\def\fig#1{fig.~#1}
\def\figs#1{figs.~#1}
\def\figalign#1#2#3#4#5#6{
\begin{figure}
\centerline{
\hbox to 2.5truein{\vtop{\hsize=2.5truein\epsfxsize=6cm
\centerline{\epsfbox{#1} }
\caption[]{#3}
\figlabel{#2}
}}
\qquad\hbox to 2.5truein{\vtop{\hsize=2.5truein\epsfxsize=6cm
\centerline{\epsfbox{#4} }
\caption[]{#6}
\figlabel{#5}
}}
}
\end{figure}
}

\def\be{\begin{equation}}
\def\ee{\end{equation}}
\def\bea{\begin{eqnarray}}
\def\eea{\end{eqnarray}}

\begin{document}
\begin{titlepage}.
\begin{flushright}
{ ~}\vskip -1in
CERN-TH/9669\\
US-FT-13/96\\
UTTG-96-03\\
\date\\
hep-th/9604004
\end{flushright}
\bigskip
\begin{center}
{\LARGE SOFTLY BROKEN $N=2$ QCD}
\vskip 0.9truecm

{Luis \'Alvarez-Gaum\'e$^{a}$,
Jacques Distler$^{b}$,\\
 Costas Kounnas$^{a}$\footnote{On leave from Dept. de Physique Th\'eorique,
ENS, 24 Rue Lhomond, F-75231, Paris, Cedex 05 France.}
 and Marcos Mari\~no$^{a,c}$}

\vspace{1pc}

{\em $^a$ Theory Division, CERN,\\
 1211 Geneva 23, Switzerland.\\
 \bigskip
  $^b$ Theory Group, Physics Department\\
  University of Texas at Austin\\
 Austin TX 78712 USA. \\
 \bigskip
$^c$ Departamento de F\'\i sica de
Part\'\i culas,\\ Universidade de Santiago
de Compostela,\\ E-15706 Santiago de Compostela, Spain.}\\

\vspace{5pc}

{\large \bf Abstract}
\end{center}

We analyze the possible soft breaking of $N=2$
supersymmetric Yang-Mills theory with and
without matter flavour preserving the analyticity properties of the
Seiberg-Witten solution. For small supersymmetry breaking
parameter with respect to the dynamical scale of the
theory we obtain an exact expression for the effective
potential.
We describe in detail the onset
of the confinement transition and some of
the patterns of chiral symmetry breaking.
If we extrapolate the results to the limit where
supersymmetry decouples, we obtains hints indicating
that perhaps a description of the QCD vacuum will
require the use of Lagrangians containing simultaneously
mutually non-local degrees of freedom (monopoles and
dyons). 

\vfill

\end{titlepage}

\def\theequation{\thesection.\arabic{equation}}

\section{Introduction}
\setcounter{equation}{0}
In two remarkable papers \cite{swone,swtwo},
 Seiberg and Witten obtained
 exact information on the dynamics of $N=2$ supersymmetric gauge
 theories in four dimensions with gauge group
$SU(2)$ and $N_{f} \le 4$
 flavour multiplets. Their work was extended to
other groups in \cite{groups}.
 One of the crucial advantages of using
$N=2$ supersymmetry is that the low-energy
 effective action in the Coulomb phase up to two derivatives
 ({\it i.e.} the K\"ahler potential,
the superpotential and the gauge kinetic function in
$N=1$ superspace language)
 are determined in terms of a single holomorphic function called the
 prepotential \cite{four}.
In references \cite{swone,swtwo},
the exact prepotential was determined using
some plausible assumptions and many consistency conditions.
For $SU(2)$ the solution is neatly presented by associating
to each case an elliptic curve together with
a meromorphic differential
of the second kind whose periods completely
determine the prepotential.
For other gauge groups \cite{groups} the solution
is again presented in terms
of the period integrals of a meromorphic differential
on a Riemann surface
whose genus is the rank of the group considered.
It was also shown in \cite{swone,swtwo} that by soft breaking $N=2$
down to $N=1$ (by adding a mass term for the adjoint $N=1$ chiral
multiplet in the $N=2$ vector multiplet) confinement follows
due to monopole condensation \cite{five}.

For $N=1$ theories exact results have also been obtained
\cite{none} using the holomorphy
properties
of the superpotential and the gauge kinetic function,
culminating in Seiberg's non-abelian
duality conjecture \cite{nadual}.

With all this new exact information it is also tempting to obtain
 exact information about ordinary QCD. The obvious problem
encountered
 is supersymmetry breaking. A useful avenue to explore is soft supersymmetry
breaking. The structure of soft supersymmetry breaking in $N=1$ theories has
been known for some time \cite{soft}. In \cite{softone,softwo}
 soft breaking terms are used to explore $N=1$ supersymmetric QCD
(SQCD)
 with gauge group $SU(N_c)$ and $N_f$ flavours of quarks,
 and to extrapolate the exact results in \cite{none}
concerning the superpotential and the phase structure
 of these theories in the absence of supersymmetry.
 This leads to expected and unexpected predictions for
 non-supersymmetric theories which may eventually be accessible to
lattice computations. In some cases however
(for instance when $N_f \ge N_c$) it is known in the supersymmetric
 case that the origin of moduli space is singular,
 and therefore some of the assumptions made about the K\"ahler
potential
 for meson and baryon operators are probably too strong.
 Since the methods of \cite{swone,swtwo} provide us with
 the effective action up to two derivatives,
 the kinetic and potential term for all low-energy fields
 are under control, and therefore in this paper we prefer
 to explore in which way we can softly break
$N=2$ SQCD directly to $N=0$ while at the same time
preserving the analyticity properties of the Seiberg-Witten solution.
 This is a very strong constraint and there is, essentially, only one
 way to accomplish this task: we make the dynamical scale
 $\Lambda$ of the $N=2$ theory a function of an
$N=2$ vector multiplet which is then frozen
to become a spurion whose $F$ and $D$-components break softly
$N=2$ down to $N=0$. If we want to interpret physically the
spurion, one can recall the string derivation of the
Seiberg-Witten solution in \cite{XI,XII} based on
type II-heterotic duality. In the field theory limit
in the heterotic side
(in order to decouple string and gravity loops)
the natural scaling is taken to be
$M {\rm e}^{iS} =\Lambda$, where $M$ is the Planck mass,
 $S$ is the dilaton (in the low-energy theory $S=\theta/2\pi+ 4\pi
i/g^2$,
 with $g$ the gauge coupling constant and
$\theta$ the CP-violating phase),
and $\Lambda$ the dynamical scale of the gauge theory
which is kept fixed while
$M \rightarrow \infty$ and $iS \rightarrow \infty$.
Since the dilaton sits in a vector multiplet of $N=2$
when the heterotic string is compactified on
$K3 \times T_{2}$, this is precisely the field
we want to make into a spurion, and we show later
that this procedure is compatible with the
Seiberg-Witten monodromies. In this way we obtain a
theory at $N=0$ with a more restricted structure
that those used in \cite{softone,softwo}.
As a consistency check, we start along the lines of
\cite{XI,XII} with the theory coupled to $N=2$ supergravity
 with a simple superpotential which breaks spontaneously
supersymmetry through an auxiliary field associated to
the graviphoton, which also gives vacuum expectation
values to the auxiliaries in the dilaton multiplet.
At low-energies one obtains a theory with all the
allowed soft breakings, however in the scaling
limit mentioned previously, the only surviving
soft terms are those one would obtain had we worked from
the beginning with the rigid $N=2$ theory plus the
dilaton spurion. As soon as the soft breaking terms
are turned on monopole condensation appears,
 and we get a unique ground state
(near the massless monopole point of
\cite{swone,swtwo}). Furthermore, in the Higgs
region we can compute
the effective potential, and we can verify that this potential
drives the theory towards the region where
condensation takes place. When the supersymmetry breaking parameter
is increased, the minimum displaces to the right along the
real $u$-axis. At the same time, the region in the $u$-plane
in which the monopole condensate is energetically-favoured expands.
Near the massless dyon point of \cite{swone,swtwo}, we find that dyon
condensation is energetically favourable but, unlike monopole condensation,
it is not sufficiently-strong an effect to lead to another minimum of the
effective potential. Eventually, when the soft supersymmetry breaking
parameter is made sufficiently large, the regions where monopole and dyon
condensation are favoured begin to overlap. At this point, it is clear that
our methods break down, and new physics is needed to describe the dynamics of
these mutually-nonlocal degrees of freedom.

  One advantage of this method
of using the dilaton spurion to softly break supersymmetry
from $N=2$ to $N=0$ is its universality.
It works for any gauge group and any number
of massive or massless quarks. As a further example
we consider the theory with two hypermultiplets
of massless quarks. The global symmetry is
$O(4)\times SU(2)_R \times U(1)_R$,
where $SU(2)_R$ is the $R$-symmetry associated to
$N=2$ supersymmetry. Monopole condensation leads
to a peculiar pattern of chiral symmetry breaking.
 Writing $SO(4)=SU(2)_l\times  SU(2)_r$,
we find that near the massless monopole region
$SU(2)_r$ breaks completely while $SU(2)_l$ remains intact.
Due to the properties of the $N=2$ solution in
\cite{swone,swtwo} we can compute the low-energy
Goldstone boson Lagrangian reliably at least
for small supersymmetry breaking parameter.
We also find two Higgs branches corresponding to the two
Higgs phases described in \cite{swtwo}. As one would
expect, they are
smoothly connected to the confining phase.

The organization of this paper is as follows:
In section two we collect some useful formul\ae\ summarizing
 the main features on \cite{swone,swtwo} which are needed in
later sections. In section three we analyze the
 effective action once
the dilaton spurion is included. The modular transformations
 of the action and coupling constants will be derived,
agreeing with the general results derived in \cite{XIII}
concerning the modification of the symplectic
transformations of special geometry in the
presence of background $N=2$ vector superfields.
 There are some interesting consequences of the
modular transformations related to the fact that
 in the moduli space of the $N=2$ theory we have
 to use different effective actions in different
patches such that the light fields in different
patches are not mutually local. In section four
we derive the same action starting with the
$N=2$ supergravity theory and spontaneous breaking
 of supersymmetry. Section five presents the detailed
 analysis of the low-energy effective action, the
onset of monopole condensation and the numerical
results. In section six we extend our results to the case of
$SU(2)$ with two massless quark hypermultiplets.
Finally in section seven we present the conclusions and outlook.

\section{The Seiberg-Witten Solution}
\setcounter{equation}{0}

We will concentrate for simplicity on the case of
$SU(2)$ with $N_f=0,2$ flavours of quarks.
 Because of the different normalization of the charge generator
 in \cite{swone} and \cite{swtwo} due to the presence
 of flavours, the elliptic curve in these two cases
 is the same, and most of the analytic and numerical
 computations are exactly the same. In the $N_f=0$
case the classical theory is described by a quadratic prepotential
\be
\label{one}
{\cal F}^{\rm cl}= {1 \over 2}\tau^{\rm cl} (A^a)^2 \\
\ee
\be
\tau^{\rm cl}={\theta \over 2\pi}+{4 \pi i \over g^2}
\label{ii}
\ee
where $A^a$, $a=1,2,3$,
are the $N=2$ vector multiplets associated
 to the generators of $SU(2)$.
In terms of $N=1$ multiplets $A^a$
contains a vector multiplet $(A^a_{\mu}, \lambda^{a})$, and a chiral
multiplet $(\psi^a, \phi^a)$.
Hence it describes a vector,
two Majorana fermions and a complex scalar;
all in the adjoint representation.
$N=2$ supersymmetry does not allow
 a superpotential for the theory and
therefore the scalar potential
 is purely $D$-term:
\be
V(\phi)={1 \over g^2} {\rm Tr} [\phi, \phi^{\dagger}]^2
\label{iii}
\ee
There is a moduli space of vacua.
The minima of (\ref{iii})
 can be taken to be of the form $\phi ={1\over 2}a\sigma^3$
 with $a$ complex. A gauge invariant description of this
 moduli space is provided by the variable
$u ={\rm Tr}\phi^2={1\over 2}a^2$ at the classical level.
 Each point in this moduli space represents a different theory.
 For $a \not= 0$ the charged multiplets acquire a mass
$M=\sqrt {2}|a|$, and $SU(2)$ is spontaneously broken to $U(1)$,
 and at $a=0$ the full $SU(2)$ symmetry is restored.
Away from the origin we can integrate out the massive
multiplets and obtain a low-energy effective theory
which depends only on the ``photon" multiplet.
The theory is fully described in terms of a prepotential
${\cal F}(A)$. The lagrangian in $N=1$ superspace is
\be
{\cal L}={1 \over 4 \pi}{\rm Im} \Big[ \int d^4 \theta {\partial
F\over \partial A} {\overline A} + {1\over 2} \int d^2 \theta
{\partial ^2 F \over \partial A^2} W_{\alpha}W^{\alpha} \Big].
\label{iv}
\ee
The K\"ahler potential and gauge kinetic functions are given in
general by:
$$
K(a, {\bar a})={1 \over 4 \pi}{\rm Im} a_{D,i}{\bar a}^i,
$$
$$
\tau_{ij}={1\over 2} {\partial ^2 {\cal F} \over \partial a^i
\partial a^j}, $$
\be
a_{D,i} \equiv {\partial {\cal F} \over \partial a^i}.
\label{v}
\ee
In perturbation theory ${\cal F}$ only receives
one-loop contributions.
 The important thing is to determine the
non-perturbative corrections.
This was done in \cite{swone,swtwo}. Some of the properties
of the exact solution are:

i) The $SU(2)$ symmetry is never restored. The theory stays
in the Coulomb phase throughout the $u$-plane.

ii) The moduli space has a symmetry $u \rightarrow -u$
 (the non-anomalous subset of the $U(1)_R$ group),
and at the points $u=\Lambda^2$, $-\Lambda^2$
singularities in ${\cal F}$ develop.
Physically they correspond respectively to a massless
monopole and dyon with charges $(q_e,q_m)=(0,1)$, $(-1,1)$.
Hence near $u=\Lambda^2$, $-\Lambda^2$ the correct
effective action should include together
with the photon vector multiplet monopole or dyon hypermultiplets.

iii) The function ${\cal F}(a)$ is holomorphic.
It is better to think in terms of the vector
$^{t}v=(a_D, a)$ which defines a flat $SL_2({\bf Z})$
vector bundle over the moduli space ${\cal M}_u$
(the $u$-plane).
Its properties are determined by the singularities and
the monodromies around them. Since
${\partial ^2 {\cal F} / \partial a^2}$ or
${\partial a_D/ \partial a}$ is the coupling constant,
these data are obtained from the $\beta$-function in
the three patches: large-$u$, the Higgs phase,
the monopole and the dyon regions.
{}From the BPS mass formula \cite{XIV,XV}
the mass of a BPS state of charge $(q_e,q_m)$
(with $q_e$, $q_m$ coprime for the charge to be stable) is:
\be
M={\sqrt 2}|q_e a+q_ma_D|.
\label{vi}
\ee
If at some point $u_0$ in ${\cal M}_u$, $M(u_0)=0$, the monodromy
around this point is given by \cite{swone,swtwo,groups}
\be
 \left(\begin{array}{c}a_{D} \\
                    a \end{array}\right) \rightarrow M(q_e, q_m)
\left(\begin{array}{c}a_D\\
               a \end{array}\right),
\label{vii}
\ee
\be
M(q_e, q_m)=\left(\begin{array}{cc} 1+2q_eq_m & 2q_e^2 \\
                           -2q_m^2&1-2q_eq_m \end{array} \right).
\label{viii}
\ee
Also for large $u$, ${\cal F}$ is dominated by the
perturbative one loop contribution, obtained from the
one loop $\beta$-function:
\be
{\cal F}_{\rm 1- loop}(a)={i \over 2\pi} a^2{\rm ln}{a^2 \over
\Lambda}
\label{ix}
\ee
Hence we also have monodromy at infinity.
The three generators of the monodromy are therefore:
\be
M_{\infty}= \left(\begin{array}{cc}-1& 2\\
               0&-1 \end{array} \right),\,\,\,\
M_{\Lambda^2}=\left(\begin{array}{cc}1& 0\\
                                 -2&1 \end{array}\right),
 \,\,\,\ M_{-\Lambda^2}=\left(\begin{array}{cc}-1& 2\\
                                                                -2&3
\end{array}\right);
\label{x}
\ee
and they satisfy:
\be
M_{\infty}=M_{\Lambda^2}M_{-\Lambda^2}.
\label{xi}
\ee
These matrices generate the subgroup
$\Gamma_2 \subset SL_2({\bf Z})$ of $2 \times 2$
matrices congruent to the unit matrix modulo $2$.

We learn from (\ref{vi})-(\ref{vii}) that in the Higgs,
monopole and dyon patches, the natural independent
variables to use are respectively $a^{(h)}=a$, $a^{(m)}=a_D$,
$a^{(d)}=a_D-a$. Thus in each patch we have a
different prepotential:
\be
{\cal F}^{(h)}(a), \,\,\,\ {\cal F}^{(m)}(a^m), \,\,\,\ {\cal
F}^{(d)}(a^d).
\label{xii}
\ee

iv) The explicit form of $a(u)$, $a_D(u)$ is given in
terms of the
 periods of a meromorphic differential of the second
kind on a genus
 one surface described by the equation:
\be
y^2=(x^2-\Lambda^4)(x-u),
\label{xiii}
\ee
describing the double covering of the plane branched at
${\pm \Lambda^2}$, $u$, $\infty$.
We choose the cuts $\{-\Lambda^2, \Lambda^2\}$, $\{u, \infty\}$.
 The correctly normalized meromorphic 1-form is:
\be
\lambda=\Lambda{{\sqrt 2} \over 2\pi} {dx {\sqrt {x-u/\Lambda^2}}
 \over {\sqrt {x^2-1}}}.
\label{xiv}
\ee
Then:
\be
a(u)= \Lambda{{\sqrt 2} \over \pi} \int_{-1}^{1}{dt {\sqrt
{u/\Lambda^2-t}} \over {\sqrt {1-t^2}}};
\label{xv}
\ee
\be
a_D(u)= \Lambda{{\sqrt 2} \over \pi} \int_{1}^{u/\Lambda^2}{dt {\sqrt
{u/\Lambda^2-t}} \over {\sqrt {1-t^2}}}.
\label{xvi}
\ee
Using the hypergeometric representation of the elliptic functions
\cite{XVI}:
$$
K(k)={\pi \over 2}F(1/2, 1/2, 1;k^2); \,\,\ K'(k)=K(k');
$$
\be
\label{xvii}
E(k)={\pi \over 2}F(-1/2, 1/2, 1;k^2); \,\,\ E'(k)=E(k'), \,\,\
{k'}^2+k^2=1,
\ee
we obtain :
\be
k^2={2 \over 1+u/\Lambda^2}, \,\,\,\ {k'}^2={u-\Lambda^2 \over
u+\Lambda^2},
\label{xviii}
\ee
\be
a(u)={4 \Lambda \over \pi k}E(k), \,\,\,\,\,\ a_D(u)={4 \Lambda \over
i \pi }{E'(k)-K'(k) \over k}.
\label{xix}
\ee
Using the elliptic function identities:
\be
{dE \over dk}={E-K \over k}, \,\,\,\,\ {dK \over dk}={1 \over k
{k'}^2}(E-{k'}^2 K),
\label{xx}
\ee
\be
{dE' \over dk}=-{k \over {k'}^2}(E'-K'), \,\,\,\,\ {dK' \over
dk}=-{1\over k {k'}^2}(E'-{k}^2K'),
\label{xxi}
\ee
the coupling constant becomes:
\be
\tau_{11}={\partial a_{D} \over \partial a}={da_{D}/dk \over
da/dk}={iK' \over K},
\label{xxii}
\ee
which is indeed the period matrix of the curve (\ref{xiii}).

Finally, to determine the prepotential ${\cal F}={\cal F}(a)$,
we have to invert $a=a(u)$, to write $u=u(a)$,
and then integrate $a_D=\partial {\cal F}/ \partial a$.

Before closing this section, we derive the modular
transformation properties of ${\cal F}(a)$.
If $\Gamma \in SL_2({\bf Z})$,
$\Gamma=\left(\begin{array}{c@{\hspace{1 mm}}c}\alpha &\beta\\ [-.3 mm]
                           \gamma&\delta \end{array}\right)$, then
$a_D^{\Gamma}=\alpha a_D+\beta a$, $a^{\Gamma}=\gamma a_D +\delta a$.
 We want to express ${\cal F}_{\Gamma}(a^{\Gamma})$
in terms of ${\cal F}(a)$. Since
\bea
{\partial {\cal F}_{\Gamma}(a^{\Gamma}) \over \partial a} &=&
{\partial a^{\Gamma} \over \partial a}
{\partial {\cal F}_{\Gamma}(a^{\Gamma})\over \partial a^{\Gamma}}=
\Bigl( \gamma {\partial a_D \over \partial a}+ \delta \Bigr)
a_D^{\Gamma} \nonumber \\
& \nonumber \\
&=& \big( \alpha {\cal F}'+\beta \big) \big( \gamma {\cal F}''+\delta
\big), \label{xxiii}
\eea
using $\alpha \delta - \beta \gamma =1$ we obtain:
\be
{\cal F}_{\Gamma}(a_{\Gamma})=
{1\over 2} \beta \delta a^2 + {1\over 2}\alpha \gamma a_D^2+ \beta
\gamma a a_D+{\cal F}(a).
\label{xxiv}
\ee
In particular,
under the two generators $T$, $S$ of $SL_2({\bf Z})$:
\bea
&\Gamma =T= \left(\matrix{1& 1\cr
               0&1\cr}\right),\,\,\,\
{\cal F}_{T}(a_{T})={1\over 2}a^2+{\cal F}(a), \nonumber \\
&\Gamma =S= \left(\matrix{0& 1\cr
               -1&0\cr}\right),\,\,\,\
{\cal F}_{S}(a_S)=-a a_D+{\cal F}(a).
\label{xxv}
\eea

When there are flavours similar results apply
in the Coulomb phase \cite{swone,swtwo},
the solution of the model is presented in terms
 of an elliptic curve and $a(u)$, $a_D(u)$
are given by period integrals.
We will recall some details in section 6.

\section{Breaking $N=2$ with a Dilaton Spurion}
\setcounter{equation}{0}

We now would like to break $N=2$ supersymmetry preserving
 the holomorphy properties of the Seiberg-Witten solution.
In the theory without flavours we want to introduce another
$N=2$ vector multiplet $s$ in the prepotential ${\cal F}(a,s)$
 in such a way that $s$, $s_D={\partial {\cal F} / \partial s}$
 be monodromy invariant. We can then freeze the scalar and auxiliary
components of this superfield to be constants to generate soft
breaking of $N=2$. Since the only free parameter in the
Seiberg-Witten solution is $\Lambda$,
the simplest choice is to make $\Lambda$ a function of
a background vector superfield.
The scale $\Lambda$ is related to the coupling constant and
$\theta$-parameter by
$\Lambda^4\sim {\rm exp}(-{8 \pi \over g^2}+i\theta)$,
it is then natural to include a dilaton field
$S$ such that $\Lambda \sim {\rm e}^{iS}$,
${\rm Im}\,\ S \sim 1/g^2$, ${\rm Re}\,\ S \sim \theta$.
This is the correct choice if we think of the
embedding of the $N=2$ $SU(2)$ theory in the heterotic
string compactified on $K3 \times T_2$ \cite{XI,XII}
where the dilaton is part of a vector multiplet.
If we can show that ${\partial {\cal F} / \partial s}$
is invariant under the Seiberg-Witten monodromy,
the addition of this extra superfield does not
change any of the holomorphic properties of
the solution presented in section 2.
In each region of the moduli space we can
write a prepotential adapted to the local
coordinates of the form:
\be
{\cal F}=a^2f(a/\Lambda).
\label{threei}
\ee
A simple consequence of the modular transformation properties of $a$,
$a_D$ and ${\cal F}$ (\ref{xxiv}) imply that
\be
{\cal F}-{1 \over 2}a a_D
\label{threeii}
\ee
is modular invariant. Hence (\ref{threeii}) is only a
function of the moduli. To determine this function it
 suffices to note that the periods $a_D(u)$, $a(u)$
satisfy a second order differential equation
(they are hypergeometric functions),
the Picard-Fuchs equation for the curve (\ref{xiii}):
\be
{d^2 \omega \over d u^2} + {1 \over u^2-\Lambda^4}\omega=0.
\label{threeiii}
\ee
The absence of a first derivative term in (\ref{threeiii})
implies that the Wronskian of the two independent
solutions $a_D(u)$, $a(u)$: $a da_D/du- a_D da/du$
is a constant, whose value can be determined by
evaluating it in the weak coupling (large $u$) region.
 Integrating the wronskian with respect to $u$ leads to
\be
{\cal F}-{1 \over 2}a a_D =-{i\over \pi}u.
\label{threeiv}
\ee
This relation was first derived in \cite{matone}
and further explored in \cite{XVIII}.
Once the spurion field $S$ is introduced,
we have two vector multiplets $a^0 \equiv s$,
$a^1\equiv a$, and a $2\times 2$ matrix of couplings:
\be
\tau_{11}={\partial^2 {\cal F} \over \partial^2 a},\,\,\,\
\tau_{01}={\partial^2 {\cal F} \over \partial s \partial a},\,\,\,\
\tau_{00}={\partial^2 {\cal F} \over \partial ^2 s},
\label{threev}
\ee
whose modular properties and explicit representation
 we would like to determine. From (\ref{threei})
plus the identification $\Lambda={\rm e}^{iS}$,
we obtain ($\partial / \partial s =i\Lambda
\partial /\partial \Lambda$):
\bea
a_D&= 2af+{a^2\over \Lambda}f', \,\,\,\,\ \tau_{11}=2f+4{a\over\Lambda}
f'+{a^2 \over \Lambda^2}f'',\nonumber\\
\tau_{01}&=-{3ia^2 \over \Lambda}f'-{ia^3 \over \Lambda^2}f'',
\,\,\,\,\ \tau_{00}=-{a^3 \over \Lambda}f'-{a^4 \over \Lambda^2} f'';
\label{threevi}
\eea
and from (\ref{threevi}) we obtain
\bea
\tau_{01}&=&i(a_D-a\tau_{11}), \,\,\,\,\,\,\,\,\,\,\,\ {\partial
\tau_{01}\over \partial a}=-ia{\partial \tau_{11} \over \partial
a};\nonumber\\
{\partial \tau_{00}\over \partial a}&=&i\tau_{01}-a^2{\partial
\tau_{11}\over \partial a}.
\label{threevii}
\eea
In particular:
\be
{\partial {\cal F} \over \partial  s}=2i\Big({\cal F}-{1 \over 2}a
a_D\Big)= {2 \over \pi} u
\label{threeviii}
\ee
and:
$$
\tau_{01}={2 \over \pi}{\partial  u \over \partial a},
$$
\be
 \tau_{00}={2i \over \pi}\Big(2 u-a{\partial u \over \partial
a}\Big).
\label{threeix}
\ee
The last equation in (\ref{threeix}) is obtained by
integrating ${\partial \tau_{00}/ \partial a}$
using (\ref{threeviii}). A lesson we draw from (\ref{threeviii})
is the monodromy invariance of
$s_D={\partial {\cal F} / \partial  s}$,
although we will obtain this result from a
 more indirect procedure later. Finally in
writing $\tau_{00}$ in (\ref{threeix}) we have
set to zero an integration constant depending
only on $s$. This is the result we would have obtained
 had we started with the Seiberg-Witten solution and
 compute $\tau_{00}$ as
$-(\Lambda \partial / \partial \Lambda)^2{\cal F}$.
 As an application of (\ref{threevii})-(\ref{threeix})
we can compute the couplings $\tau_{ij}$
in the Higgs and monopole region.

i) Higgs region:

$$
a_D^{(h)}={4 \Lambda \over i\pi}{E'-K' \over k},
\,\,\,\,\ a^{(h)}={4 \Lambda \over \pi k}E(k),
$$
\be
\tau^{(h)}_{11}={iK' \over K},\,\,\,\ \tau^{(h)}_{01}=
{2\Lambda \over kK},\,\,\,\ \tau^{(h)}_{00}=
-{8i \Lambda^2\over \pi} \Big( {E-K \over k^2K} + {1 \over 2}\Big).
\label{threex}
\ee

ii) Monopole region:

$$
a_D^{(m)}={4 \Lambda \over \pi k}E(k),
\,\,\,\,\ a^{(m)}=-{4 \Lambda \over i\pi}{E'-K' \over k},
$$
\be
\tau^{(m)}_{11}={iK \over K'},\,\,\,\
\tau^{(m)}_{01}={2i\Lambda \over kK'},\,\,\,\
\tau^{(m)}_{00}={8i \Lambda^2\over \pi}
\Big( {E'\over k^2K'} -{1 \over 2}\Big).
\label{threexi}
\ee
Between (\ref{threex}) and (\ref{threexi})
we find an apparent puzzle. If we compute the
 difference between $\tau^{(m)}_{00}$ and
$\tau^{(h)}_{00}$ the result is not zero
as one might na\"\i vely expect:
\be
\tau^{(m)}_{00}-\tau^{(h)}_{00}={4i\Lambda^2 \over k^2 KK'}.
\label{threexii}
\ee
Before we showed that ${\partial {\cal F}/\partial  s}$
is a monodromy invariant, thus one would be tempted
to believe that ${\partial^2 {\cal F} /\partial  s^2}$
is also invariant and that it should take the same
values in the Higgs and monopole region.
The reason for this apparent mismatch has
to do with the fact that the light fields
in the two regions are not mutually local,
and ${\cal F}$ is written in each region in
terms of the light fields. We can compute the
difference (\ref{threexii}) on general grounds
 as follows. In a region where the coordinate
describing the light fields is $a_{\Gamma}$
($\Gamma$ an element of $SL_2({\bf Z})$), the prepotential is:
\be
{\cal F}_{\Gamma}={\cal F}_{\Gamma}(a_{\Gamma}, s)=
a_{\Gamma}^2f_{\Gamma}(a_{\Gamma}/\Lambda),
\label{threexiii}
\ee
with couplings:
\be
\tau_{ij}^{\Gamma}= {\partial^2
{\cal F}_{\Gamma} \over \partial a_{\Gamma}^i \partial a_{\Gamma}^j}.
\label{threexiv}
\ee
As $a_{\Gamma} =a_{\Gamma}(a,s)$,
 we must be careful in computing the derivatives
(as in Thermodynamics). This will give us the
transformation rules of $\tau_{ij}^{\Gamma}$. Since $a_{\Gamma}
=a_{\Gamma}(a,s)=\gamma a_D(a,s) +\delta a$,
$\Gamma=\left(\begin{array}{cc}\alpha&\beta\\
                           \gamma&\delta\end{array}\right)$,
 we have:
\be
\left(
      \begin{array}{cc} {\displaystyle {\partial a_{\Gamma} \over \partial
a}}& {\displaystyle {\partial
a_{\Gamma}
\over \partial s} }\\
& \\
{\displaystyle {\partial s \over  \partial a }} & {\displaystyle {\partial s
\over  \partial s}}
\end{array}\right) =
\left(\begin{array}{cc}{\gamma \tau_{11}+\delta}&{\gamma \tau_{01}}\\
                           0&1\end{array}\right),
\label{threexv}
\ee
with inverse
\be
\label{threexvi}
\left( \begin{array}{cc} {\displaystyle{\partial a \over \partial
a_{\Gamma}}} & {\displaystyle {\partial
a \over \partial s }}  \\
 &  \\
     {\displaystyle {\partial a \over \partial a_{\Gamma}}} &
{\displaystyle {\partial s \over \partial s}}
\end{array}\right) =
{1 \over \gamma \tau_{11}+\delta } \left( \begin{array}{cc}1&-{\gamma
\tau_{01}}\\
                           0&{\gamma
\tau_{11}+\delta}\end{array}\right).
\ee
In particular,
$$
\Bigl( {\partial \over \partial a_{\Gamma}} \Big)_{\Gamma -{\rm
basis}}
={1 \over \gamma \tau_{11}+\delta }{\partial \over \partial a},
$$
\be
\Big( {\partial \over \partial s} \Big)_{\Gamma -{\rm basis}}
={\partial \over \partial s}-{\gamma \tau_{01} \over  \gamma
\tau_{11}+\delta}{\partial \over \partial a};
\label{threexvii}
\ee
and together with the transformation rules for ${\cal F}_{\Gamma}$
(\ref{xxiv}), (\ref{threexvii}) leads to:
\bea
\tau_{11}^{\Gamma}&=&{\alpha \tau_{11}+ \beta \over \gamma
\tau_{11}+\delta},
 \,\,\,\,\,\,\,\,\,\ \tau_{01}^{\Gamma}={\tau_{01} \over \gamma
\tau_{11}+\delta },\nonumber \\
\tau_{00} ^{\Gamma}&=&\tau_{00}-{\gamma \tau_{01}^2 \over \gamma
\tau_{11}+\delta}.
\label{threexviii}
\eea
The $\Gamma$-transformations which change $\tau_{00}$
 are those for which $\gamma \not= 0$,
but these are precisely the ones mixing
non-trivially the electric and magnetic fields.
 With the explicit formul\ae\ (\ref{threex}) and
(\ref{threexi}) it is easy to verify that
(\ref{threexii}) follows from (\ref{threexviii}).
Furthermore, to check that ${\partial {\cal F}/\partial  s}$
 is modular invariant it suffices to prove that
$({\partial {\cal F}_{\Gamma}/\partial  s})_{{\Gamma}-{\rm
basis}}={\partial {\cal F}/\partial  s}$; a straightforward
consequence of the previous equations.
Similarly, but with some more algebra, one can verify:
\be
K_{\Gamma}={\rm Im}A_{D,i}^{\Gamma}{\overline A}^{\Gamma i}= {\rm
Im}\Bigl({\partial {\cal F}_{\Gamma}\over \partial  s}{\Big
|}_{{\Gamma}-{\rm basis}}{\bar s}+{\partial {\cal F}_{\Gamma}\over
\partial  a_{\Gamma}}{\Big |}_{{\Gamma}-{\rm basis}}{\bar
a}^{\Gamma}\Bigr)=K(A,S).
\label{threexix}
\ee
An illuminating way to obtain the transformation
(\ref{threexviii}) when $\Gamma=S= \left(\begin{array}{cc}0& 1\\
               -1&0\end{array}\right)$
is to start with the $N=1$ superspace action:
\be
{1 \over 4\pi}{\rm Im} \int \big({1 \over 2}
\tau_{11}W_1 W_1 + \tau_{01} W_0W_1 +{1 \over 2}\tau_{00}W_0W_0
\big).
\label{threexx}
\ee
$S$-duality follows by adding
\be
{1 \over 4\pi}{\rm Im} \int W_D W_1
\label{threexxi}
\ee
to (\ref{threexx}) and integrating out $W_1$.
This yields the dual action
\be
{1 \over 4\pi}{\rm Im} \int \Bigl(-{1 \over 2\tau_{11}}W_D W_D +
{\tau_{01} \over\tau_{11}} W_0W_D +{1 \over
2}\big(\tau_{00}-{\tau_{01}^2 \over \tau_{11}} \big) W_0W_0 \Bigr),
\label{threexxii}
\ee
in exact agreement with (\ref{threexviii}). These transformation
rules also agree with the general formul\ae\ in \cite{XIII}.

Now we have all the ingredients to write the
low-energy effective action including the spurion.
To analyze the vacuum structure we also need to
 include in the monopole (and dyon) region the
coupling to the
 monopole hypermultiplets. In rigid $N=2$ supersymmetry
the scalar
 components of a hypermultiplet take values in a hyperk\"ahler
manifold \cite{XIX}. If we denote by $m$, $\widetilde m$ the complex
scalar
 components of the monopole multiplet, the $SU(2)_R$-symmetry of
$N=2$ supersymmetry implies that $(m, \widetilde m)$ form a doublet
under this symmetry. $(m, \widetilde m)$ have opposite $U(1)$ charges.
 Hence the hyperk\"ahler manifold has complex dimension two and
must have an isometry group $SU(2) \times U(1)$.
If we knew some properties of the theory for large values of
$(m, \widetilde m)$ we could determine the asymptotic structure
 of the monopole manifold. Assuming no global identifications
 at large values of $m$, $\widetilde m$, the only
two natural choices would be flat space and the Taub-Nut instanton.
 In four dimensions hyperk\"ahler manifolds are equivalent
 to gravitational instantons with self-dual connections.
With the given isometry group we can identify flat space,
Eguchi-Hanson and Taub-Nut. However in the Eguchi-Hanson
instanton the space is asymptotically $S^3/{\bf Z}_2$,
and in the Taub-Nut case it looks asymptotically like
$S^3$ but in a distorted form: it is given by the Hopf
fibration of $S^3$ over $S^2$, where the $S^1$-fibre
reaches a constant asymptotic value whereas the radius
of the $S^2$-base goes to infinity. It does not seem
physically reasonable to impose such behaviour for
large monopole fields. However one should not extrapolate
 the effective action to that region. We will assume that
 the hyperk\"ahler manifold is ${\bf C}^2$. For small fields
 this is a good approximation.
Since the monopoles come in a hypermultiplet,
 in a heterotic string they do not couple to the
dilaton in the first two terms in the effective action.
 Therefore the monopole Lagrangian will be taken to be:
\be
{\cal L}_M=\int d^4 \theta \big( M^{*}{\rm e}^{2V_D}M +
 {\widetilde M}^{*}{\rm e}^{-2V_D}{\widetilde M}\big)+ \Bigl( \int
d^2 \theta {\sqrt 2}A_DM {\widetilde M} + {\rm h.c.} \Big)
\label{threexxiii}
\ee
where $A_D$ is the chiral multiplet in the $N=2$
vector multiplet of
the dual photon \cite{swone,swtwo}. Its scalar
component is $a_D$,
a good coordinate in the $u=\Lambda^2$ region of
the moduli space
where the monopole becomes massless.
The full lagrangian is given by
adding up (\ref{iv}) and (\ref{threexxiii}).
Here we should be careful with the
prepotential ${\cal F}(A,S)$ that is included in (\ref{iv}).
The exact solution (\ref{xv}), (\ref{xvi}),
(\ref{xxii}) describes the  Wilsonian effective
action where all states but the photon multiplet
are integrated out,
 in particular the monopoles. Near $u=\Lambda^2$, where the
monopole becomes massless in the $N=2$ theory, we have to include
 (\ref{threexxiii}) in the effective action and we should
 be careful in not overcounting the monopole contribution
 in ${\cal F}(A)$.

 We have already integrated out the quantum fluctuations of the monopole;
they are already represented in (\ref{iv}). What appears in
(\ref{threexxiii}) is the {\it classical} monopole field. In order to find
the vacuum, we still need to extremize with respect to it. In fact, as
Lorentz-invariance is unbroken, we really need only concern ourselves with
the constant mode of the monopole field.  Our task, then will be to minimize
the effective potential with respect to the classical monopole field.

One way to think about this is that, in obtaining the Wilsonian effective
action (\ref{iv}) at low energies, we have integrated out all of the
nonzero-momentum modes of the monopole field, but we have not (yet)
integrated out the constant mode. Since, in the softly-broken case (as we
shall see) all of the scalars are massive, there is, essentially, no
difference between the Wilsonian and 1PI effective actions. The latter, for
the constant modes of the fields is just the usual effective potential,
 $V_{\rm eff}$ \cite{pc}.

What we will find is that, over most of the $u$-plane,
including
 the monopole
has no effect on $V_{\rm eff}(u)$. The extremum occurs at zero monopole VEV.
However, there will be a region, near $u=\Lambda^2$,
where a nonzero monopole VEV is favoured and the effect of including
(\ref{threexxiii}) is to {\it lower} the energy.

Therefore, to determine the vacuum structure in this region, we {\it must}
add up (\ref{threexxiii}) with 
$$
{\cal L}={1 \over 4 \pi}{\rm Im}
\Bigl[ \int d^4 \theta {\partial F\over
\partial A^i} {\overline A}^i + {1\over 2}
\int d^2 \theta {\partial ^2 F \over
\partial A^i \partial A^j} W^i_{\alpha}W^{\alpha  j} \Bigr],
$$
\be i=0,1; \,\,\,\ A^0=S, \,\,\ A^1=A,
\label{threexxiv}
\ee
using the complete prepotential in the Seiberg-Witten solution.
We read off the potential by keeping non-derivative terms
and auxiliary fields. $S$ is frozen to be a constant.
Its lowest component fixes the scale $\Lambda$ but we
also freeze its auxiliaries $F_0$, $D_0$
(from the chiral and the $N=1$ vector multiplets,
respectively). Eliminating the auxiliary fields $F_m$,
 $F_{\widetilde m}$ and $F_a$ we obtain a potential:
\bea
V &=& {1 \over 2b_{11}}\big( |m|^2+|\widetilde m|^2 \big)^2 + 2|a|^2 \big(
|m|^2+|\widetilde m|^2 \big) \nonumber \\
& & \nonumber\\
 &+& {1 \over b_{11}}\Bigl( {\sqrt 2}b_{01}
\big({\overline F_0}m\widetilde m + F_0{\overline m}
{\overline{\widetilde m}} \big)+b_{01}D_0
\big( |m|^2-|\widetilde m|^2 \big)\Bigr) \nonumber \\
& & \nonumber \\
&-& {{\rm det} \,\ b_{ij} \over b_{11}}
\Bigl({1\over 2}D_0^2+|F_0|^2\Bigr),
\label{threexxv}
\eea
where
\be
b_{ij}\equiv {1 \over 4 \pi}{\rm Im} \,\ \tau_{ij}={1 \over 4
\pi}{\rm Im} {\partial^2 {\cal F} \over \partial a^i \partial a^j}.
\label{threexxvi}
\ee
$m$, $\widetilde m$ are, as before, the scalar components of $M$,
 $\widetilde M$; in the same way $a$ is
taken as the scalar component of $A$,
and ${\cal F}$ is the exact solution of Seiberg and Witten.
For small values of $F_0$, $D_0$ with respect to
$\Lambda$ (\ref{threexxv}) is the exact expression including
supersymmetry breaking.
 Note that in (\ref{threexxv}) not all allowed
soft breaking terms from the $N=1$ point of view appear.
We do not have for instance a diagonal mass for $m$, $\widetilde m$,
 $B(|m|^2+|\widetilde m|^2)$, or the trilinear term
$A(am\widetilde m+{\bar a}{\overline m}{\overline{\widetilde m}})$,
but we have a $\mu$-term $\sim m\widetilde m + {\rm c.c.}$
and a cosmological term. If we look at the fermion terms
 there are also gluino masses induced, for both
sets of spinors associated to the vector multiplet.
The terms in $V$ that remain after $D_0, F_0 \rightarrow 0$
 are $SU(2)_R$ invariant as expected. More important,
$V$ contains the contribution for the metric coming
from the K\"ahler potential. This information is missing
when we only consider soft breaking in $N=1$ theories
where one may hope to control the superpotential but not
the kinetic terms. This is an important advantage of
starting with $N=2$ SQCD, the disadvantage is the
presence of an extra adjoint chiral multiplet.
Using the
monodromy transformations of the couplings (\ref{threexviii}) one can see that
${{\rm det} \,\ b_{ij}/ b_{11}}$ is a monodromy
invariant. To prove it, it is sufficient to check the invariance under
the generators $S$, $T$ of the modular group. Under $T$ it is obvious,
and for $S$ it can be done with a little algebra. This tells us
that in the vacuum energy we are taking into
account the quantum fluctuations in the right
way for different patches.

In section five we analyze in detail the potential
(\ref{threexxv}).
In the next section we derive the same action (\ref{threexxiii})
plus (\ref{threexxiv}) starting from the spontaneously broken theory
 coupled to $N=2$ supergravity. The same set of soft breaking
terms is obtained in the flat limit, including
the cosmological term. This reassures us that
we are not missing any important term. The reader not
 interested in this derivation can skip directly to section five.

\section{A Brief Foray into $N=2$ Supergravity}
\setcounter{equation}{0}
In order to give a physical meaning to the soft breaking terms it
is necessary to justify their origin in a more fundamental theory in
which the $N=2$ supersymmetry is spontaneously broken with zero (or
almost zero) cosmological term. This requirement implies that
supersymmetry must be local and that the two gravitini will become
massive via an $N=2$ superhiggs phenomenon \cite{four,ne2flat}.
Thus, our starting point must be an $N=2$ supergravity coupled to
($n_v+1$)-vector multiplets in which the desired superhiggs breaking
takes place
with vanishing vacuum energy at the classical level \cite{ne2flat}.
It is
interesting that
the structure of the $N=2$ supergravity theories with the above
properties are quite restricted and are based on a prepotential which
has the following
form \cite{ne2flat}:
\be
{\cal F}=\frac{1}{x_0}d_{abc}x^ax^bx^c,
\label{prepot}
\ee
where $x^a$ $a=1,2,\ldots,n_v$ are the matter vector multiplets and $x^0$
is an extra auxiliary vector multiplet in association with the
graviphoton of the $N=2$ gravitational multiplet. In this section ${\cal F}$
denotes the prepotential in $N=2$ supergravity, not to be confused with the
Seiberg-Witten prepotential.

The above choice of the prepotential defines a particular class of
K\"ahler potential of the no-scale type \cite{nsm,ne2flat}:
\be
K=-{\rm log}~Y,
\label{ka}
\ee
with
\bea
 Y &=& i(x^I{\overline {\cal F}}_I - {\bar x}^I{\cal F}_I)\nonumber\\
 &=& -i\Big(2({\cal F}-{\overline {\cal F}})-(x^a-{\bar x}^a)({\cal F}_a+
{\overline{\cal F}}_a)\Big)\nonumber\\
 &=&-id_{abc}(x^a-{\bar x}^a)(x^b-{\bar x}^b)(x^c-{\bar x}^c),
\label{kahler}
\eea
where the subscripts indicate differentiation with respect to the
corresponding variable. In the above equations we denote by
$x^I=(x^0~,x^a)$ and after the
algebraic operations we choose the gauge $x^0=1$.
The breaking of supersymmetry implies the existence of a superpotential
for the vector multiplets, $W_v(X^I)|_{x^0=1}$. The form of $W$ is
restricted
by $N=2$ supersymmetry to be a homogeneous function of degree one
in  $x^I$ \cite{ne2flat,fklz}:
\be
W=g_Ix^I-f^I{\cal F}_I.
\label{superpot}
\ee
An interesting subclass of models are those in which the
prepotential is given by:
\be
F=\frac{1}{x^0}s(z^2-y^2_i).
\label{ef}
\ee
In that case the  K\"ahler manifold has an interesting structure,
namely the scalars of the vector multiplets are coordinates of the
coset
\be
\left[{SL(2,R)\over U(1)}\right ]_{s}\times \left[{SO(2,~n_v-1)
\over {SO(2)\times SO(n_v-1)}} \right]_{z,y}.
\label{coset}
\ee
This is precisely the structure which emerges in heterotic strings
with $N=2$
spacetime supersymmetry \cite{effcl,ss2}. The $s$--field is
the string dilaton--axion
vector multiplet with a $U(1)_s$ gauge field. The other abelian gauge
symmetries are
the $U(1)_{x^0}$ associated to the graviphoton of the supergravity
multiplet
and the $U(1)_z$ of the $z$--vector multiplet. The remaining gauge
group in association with the $y^i$--vector multiplets can be a
non--abelian gauge group
at particular points of the $y^i$--moduli--space. Observe that the
$U(1)_z$ cannot have a non abelian extension at any point of the
$z$--moduli--space as soon as $y^i\ne 0$. In terms of the usual
string notation, $z$ and $y$ correspond respectively to the
$T+U$ and $T-U$ combinations. The non--abelian extension
happens in some special points of the $y^i$ moduli space, e.g. the
$SU(2)_y$ extension when $y^1=0$ and $z\ne 2e^{2i\pi /3}$.
 Working in the large $z$--regime we can avoid in string theory, as
well as in the effective field theory limit, the extension of the
$U(1)_z\times U(1)_y$ to $SU(3)$ which happens at the point
$y=0,~z=2e^{i2\pi /3}$ of the moduli space.
Thus in the large $z$--regime the only non--abelian extensions
happen for special values of $y^i=0$.

We are now in a position to define in a consistent way the
Seiberg--Witten
theory in a supergravity model where supersymmetry is spontaneously
broken.
The minimal set of vector superfields at the classical level are
$x^0$, $s$,
$z$, and $y^a, a=1,2,3$, where  $a$ is the adjoint index of
$SU(2)_y$. The remaining gauge group consists of abelian factors
$U(1)_{x^0}\times U(1)_s\times U(1)_z$. Neglecting gravitational
corrections but including perturbative and non-perturbative  gauge
corrections, the $N=2$ supergravity prepotential become:
\be
{\cal F}=\frac{sz^2}{x^0}-y^2\Phi\Big({y\over x^0},{s\over x^0}\Big).
\label{stringprep}
\ee
The justification for the above expression follows from the fact that
$U(1)_{x^0}\times U(1)_s\times U(1)_z$ does not receive corrections
in the limit where we neglect the gravitational interactions. On the other
hand, the $y^2$ part receives
perturbative and non-perturbative $SU(2)$ corrections similar to those
in global supersymmetry. Obviously, one can do much better in the
context of string theory where the gravitational corrections (at
least the perturbative ones) can be also be included
\cite{effqu,ant,infr}. For our purposes however this is not necessary
since, in the end, we will take the limit in which the
gravitational interactions are neglected, keeping only the soft breaking terms.

Concerning supersymmetry breaking, we must specify our choice for
the superpotential $W_{v}(x^I)$. Although there are several
possibilities,
our choice must be consistent with the stability of the scalar
potential at the classical level, {\it i.e.}~with the existence of a
perturbative vacuum in the large $s$-limit. One consistent choice is
when
$W_v=cx^0~|_{x^0=1}$.

Finally,  we must specify the remaining interactions among the vector
multiplets and the  monopole-dyon hypermultiplets. Using the $N=1$
language
these interactions are given in terms of an effective superpotential
$W_m$ and the usual $D$-terms. $W_m$ is restricted by $N=2$
supersymmetry to have the following form \cite{fklz}:
\be
W_m=\gamma(m_Ix^I-n^I{\cal F}_I)M {\widetilde M}
\label{monopot}
\ee
To recover the results of the global case it is necessary to choose
the  $m^I,~n_I$ coefficients  to be non--zero only when $I$ is taken
in the  $y$-direction.
The total superpotential is then
\be
W_t~=~W_v+W_m
\label{total}
\ee
The remaining interactions are given by the usual $D$-terms. The
normalization
$\gamma$ of $W_m$ is fixed by $N=2$ supersymmetry (see below).

In the spirit of references \cite{XI,XII} we would like to derive
the softly
broken action of the previous
section starting with a spontaneously broken $N=2$ supergravity
theory inspired by an $N=2$ compactification
of the heterotic string. From the geometrical
 point of view this is related to
the question of how to obtain rigid special geometry
from local special geometry \cite{XXIII}.
One problem with the prepotential
in \cite{XII} is it does not admit a
straightforward flat limit. A further
change of variables is required to go to a
system of coordinates analogous to the
Calabi-Visentini variables \cite{XXIV,XXV}.
We take a different route. Together with
the dilaton and the other multiplets
in the non-gravitational part of the theory we include the
graviphoton
 in the local prepotential. String
theory suggests to start with a prepotential of the form (\ref{stringprep}):
\be
{\cal F}=sz^2-F(y,s).
\label{fouri}
\ee
The scaling limit we will take involves writing $y=a/M$, $|z|\sim 1$
and $M{\rm e}^{iS}=\Lambda$ as $M, S \rightarrow \infty$ , where $
F(y,s)$ becomes ${ 1 \over M^2}F_{\rm SW}(a, \Lambda)$, $F_{\rm SW}$
is the Seiberg-Witten prepotential. The K\"ahler potential in local
special geometry
 is constructed from (\ref{kahler}) and (\ref{fouri}) as:
\bea
i{\rm e}^{-K}=2({\cal F}-{\overline {\cal F}})&-&(s-{\bar s})({\cal
F}_s+{\overline {\cal F}}_s)\nonumber\\
&-&(z-{\bar z})({\cal F}_z+{\overline {\cal F}}_z)-(y-{\bar y})({\cal
F}_y+{\overline {\cal F}}_y).
\label{fourii}
\eea
 We also include a contribution to $K$ coming
from the monopoles of the form:
\be
\delta K=\alpha \big(|m|^2+|\widetilde m|^2\big),
\label{fouriii}
\ee
where we will have to work out the scaling properties of $\alpha$.
Finally the simplest superpotential breaking supersymmetry
spontaneously is (\ref{total}):
\be
W_t=c+{\sqrt 2}A_D M {\widetilde M} \equiv c+w.
\label{fouriv}
\ee
In the Higgs region we would simply take the constant term. There are
more general choices
for $W$, but (\ref{fouriv}) is the simplest one. Supersymmetry
breaking is primarily done by the graviphoton sector which then
communicates it through gravity to the other sectors of the theory.
Defining the $G$-function as:
\be
G=K + {\rm ln}|W|^2,
\label{fourv}
\ee
the scalar potential, after the auxiliary fields are eliminated, is
given by:
\bea
V&={\rm e}^{G} \Bigl( G_{\bar i}(G^{-1})^{{\bar i} j}G_j -3 \Bigr)
+D{\rm -terms},\nonumber \\
G_{\bar i}&=\partial_{\bar i}G, \,\,\,\,\,\
G_{j}=\partial_{j}G,\,\,\,\,\,\
G_{{\bar i} j}=\partial_{\bar i} \partial_{j} G.
\label{fourvi}
\eea
In (\ref{fouri}) the first term in the right-hand side is much bigger
than the second; hence we expand in powers of ${1 \over M}$ (the
Planck mass):
\be
{\rm e}^{-K}= i\Sigma {Z}^2 \Biggl( 1-{1 \over  {Z}^2}
\Bigl({ F}_{s}+{\overline { F}}_{s} +
{ y{\overline { F}}_y-{\bar y}{ F}_{y}-i({ F}_{s} +
{\overline { F}}_{s}) \over \Sigma }\Bigr) \Biggr),
\label{fourvii}
\ee
with
\be
\Sigma \equiv s-{\bar s},\,\,\,\,\ Z \equiv z-{\bar z}.
\label{fourviii}
\ee
To second order in $1/Z$ we have:
$$
K=-{\rm log}i\Sigma -2 {\rm log}Z+{1 \over {\Sigma}^2}\phi (s, y),
$$
\be
\phi(s,y)={ F}_s+{\overline { F}}_s + {y{\overline {
F}}_y-{\bar y}{ F}_y-i({ F}_s+{\overline { F}}_s) \over
\Sigma}.
\label{fourix}
\ee
It is now a long and tedious algebraic computation to evaluate
(\ref{fourvi}) to leading order. The answer is:
\bea
&{1 \over i\Sigma \phi_{y {\bar y}}} \big| \partial_y W \big| ^2+
{1 \over i\Sigma \phi_{y {\bar y}}Z^2}
 \Bigl( \phi_y W {\overline {\partial_y W}}+
\phi_{\bar y} {\overline W} \partial_y W  \Bigr)\nonumber\\
& +{1 \over i\Sigma Z^2} \Bigl( {\alpha}^{-1} \big| \partial_m W \big|
^2 +
{\alpha}^{-1} \big| \partial_{\widetilde m} W \big| ^2\nonumber\\
&+\alpha|c|^2\big( |m|^2+|\widetilde m|^2 \big) + 2c(w+{\bar w})
\Bigr)\nonumber\\
&+ {|c|^2 \over i\Sigma Z^4 \phi_{y {\bar y}}}
\Bigl( \Sigma \big( \phi_{y {\bar y}} \phi_s -\phi_{y {\bar y}}
\phi_{\bar s} \big)-2\phi_{y {\bar y}} \phi +
\phi_{y} \phi_{\bar y}\nonumber\\
& + \Sigma ^2 \phi_{s{\bar s}}\phi_{y {\bar y}}-\Sigma^2 \phi_{s {\bar
y}}\phi_{y {\bar s}}+\Sigma \phi_{\bar y}\phi_{y {\bar s}}- \Sigma
\phi_{y } \phi_{{\bar s}y } \Bigr)\nonumber\\
&+ {\rm l.o.t},
\label{fourx}
\eea
a slightly unwieldy expression. The l.o.t. stand for lower order
terms in $M$. It is also important to consider the kinetic term for
$y$, $\bar y$ to correctly normalize the low-energy fields. From
(\ref{fourix}) we obtain:
\bea
\phi_s&=& F_{ss}+{1 \over \Sigma}\Big((y-{\bar y})F_{ys}-2F_s \Big)-
{1 \over \Sigma^2}\Big((y-{\bar y})(F_y+{\overline F}_y) -2F+2{\overline
 F}\Big),\nonumber\\
\phi_y&=& F_{sy}+{1 \over \Sigma}\Big({\overline  F}_y-F_y+(y-{\bar
y})F_{yy}\Big), \nonumber\\
\phi_{s {\bar y}}&=& -{1 \over \Sigma}F_{ys}-{1 \over
\Sigma^2}\Big({\overline  F}_y -F_y+(y-{\bar y}){\overline F}_{yy}\Big),
\nonumber\\
\phi_{y {\bar y}}&=& {1 \over \Sigma}({\overline F}_{yy}-F_{yy}), \nonumber\\
\phi_{s {\bar s}}&=& {1 \over \Sigma^2}\Big((y-{\bar
y})(F_{ys}-{\overline F}_{ys}) -2F_s-2{\overline  F}_s\Big) \label{fourxii}\\
&&\qquad-{2 \over \Sigma^3}\Big((y-{\bar y})(F_y+
{\overline F}_y) -2F+2{\overline  F}\Big).\nonumber
\eea
Inserting (\ref{fourxii}) into (\ref{fourx}) and keeping
 leading order terms
we obtain:
\bea
&{1 \over i\Sigma \phi_{y {\bar y}}} \big| \partial_y W \big| ^2-{c
\over i\Sigma \phi_{y {\bar y}}Z^2}\big(F_{ys}+{\overline
F}_{ys}\big)
\big(\partial_y W +{\overline \partial_y W} \big)\nonumber\\
&+{|c|^2 \over i\Sigma Z^4 \phi_{y {\bar y}}}\Bigl(({\overline
F}_{yy}-F_{yy})
(F_{ss}-{\overline F}_{ss})+(F_{ys}+{\overline F}_{ys})^2 \Bigr)\nonumber\\
& +{1 \over i\Sigma Z^2}\Bigl({\alpha}^{-1} \big| \partial_m W \big|
^2 +
{\alpha}^{-1} \big| \partial_{\widetilde m} W \big| ^2 + \alpha|c|^2\big(
|m|^2+|\widetilde m|^2 \big)\nonumber\\
&  2c(w+{\bar w}) \Bigr)+ D{\rm -terms}.
\label{fourxiii}
\eea
To determine the scaling limit we want to scale $y \sim a/M$,
hence $F\sim {1 \over M^2}$, $F_y \sim {1\over M}$,
$F_{yy}\sim 1$. From the kinetic term of $y$ we learn
 that $i\Sigma Z^2 \sim 1$. As in section 3 we define
$\tau_{ij}=\partial_{ij}^2F$ for the $a, s$ variables.
The scaling inside $F_{\rm SW}$ is then:
\be
M{\rm e}^{iS}=\Lambda ,
\label{fourxiv}\ee
with $\Lambda$ fixed. Since we want to recover the
purely supersymmetric terms in the potential,
this fixes $\alpha \sim 1/M^2$. Finally the
second and third terms in (\ref{fourxiii}) define
 the scaling behaviour of $c$:
\be
c{i\Sigma \over M}=m_{3/2}.
\label{fourxv}
\ee
$m_{3/2}$ is the gravitino mass and $\Lambda$ is fixed.
 From (\ref{xiv}) we learn that
$i\Sigma =2{\rm ln} {M\over \Lambda}$.
For $M \sim M_{\rm Pl}$, $\Lambda \sim 1 \,\ {\rm GeV}$,
$i\Sigma \sim 10^2$. The last two terms in (\ref{fourxiii}) become:
\be
 {m_{3/2}^2 \over (i\Sigma)^2}\big( |m|^2+|\widetilde m|^2 \big)+
{2 m_{3/2} \over i\Sigma}(w+{\bar w}).
\label{fourxvi}
\ee
In the formal limit $i\Sigma \rightarrow \infty$,
$M\rightarrow \infty$ with $M^2{\rm e}^{i\Sigma}=\Lambda^2$
fixed, these two terms disappear; if, however,
we take $M\sim M_{\rm Pl}$, they stay but with
very small coefficients with respect to the other
soft-breaking terms in (\ref{fourxiii}).
If we were to consider the full
potential, the higher order corrections are of
two types. First those suppressed by powers of
 $1/\Sigma$, $1/\Sigma^2$, and those
suppressed by powers of $1/M$.
The latter can be ignored, while the former can be
neglected in a first approximation. Notice that
(\ref{fourxiii}) is equivalent
to (\ref{threexxv}) in the $i\Sigma \rightarrow \infty$
limit with $F_0\sim m_{3/2}$, and similarly for $D_0$.
Although we have not presented here the
explicit computation of the $D$-terms in supergravity,
they also lead to the same term in (\ref{threexxv}).

The conclusion we draw from this computation is that
the soft-breaking terms included in (\ref{threexxv})
are precisely those which
are induced from a spontaneously broken $N=2$
supergravity theory in the flat limit, and although some
soft-breaking terms like (\ref{fourxvi}) also appear,
they are suppressed with respect to the leading order
 ones in (\ref{threexxv}).
Therefore, to analyze the vacuum structure,
(\ref{threexxv}) contains all the relevant
terms and we are not missing any essential ingredient.
 This is additional support for the procedure we are following.

\section{Vacuum structure}
\setcounter{equation}{0}
We now turn to the analysis of the potential
(\ref{threexxv}). We will make two additional
technical simplifications. The first one is to
ignore the small terms in (\ref{fourxvi}).
The second one is to set $D_0=0$. This makes
the algebraic structure simpler but the conclusions
 remain the same. In minimizing the effective
potential (\ref{threexxv}) we proceed in two stages:
 first we minimize with respect to the monopoles $m$,
${\widetilde m}$; and then we look graphically for
the minima with respect to the dual photon $a$.
The explicit formul\ae\ are those in (\ref{threexi})
for the monopole region.
\be
{\partial V \over \partial {\overline m}}={1 \over b_{11}}
\big( |m|^2+|{\widetilde m}|^2 \big)m+2|a|^2 m + {{\sqrt 2} \over
b_{11}}b_{01}F_0{\overline{\widetilde m}}=0,
\label{fivei}
\ee
\be
{\partial V \over \partial {\overline {\widetilde m}}}={1 \over
b_{11}}\big( |m|^2+|{\widetilde m}|^2 \big){\widetilde m} +2|a|^2 {\widetilde m
}+ {{\sqrt 2} \over b_{11}}b_{01}F_0{\overline m}=0.
\label{fiveii}
\ee
Multiplying (\ref{fivei}) by ${\widetilde m}$, (\ref{fiveii}) by $m$ and
subtracting we obtain:
\be
{{\sqrt 2} \over b_{11}}b_{01}F_0 \big( |m|^2-|{\widetilde m}|^2 \big)=0,
\label{iii}
\ee
hence $|m|^2=|{\widetilde m}|^2$. Writing
\be
m=\rho {\rm e}^{i\alpha}, \,\,\,\,\ {\widetilde m}= \rho  {\rm
e}^{i\beta}, \,\,\,\,\ F_{0}=f_{0}{\rm e}^{i\gamma};
\label{fiveiv}
\ee
we can fix the gauge so that $\alpha =0$, and absorb $\gamma$ in
$\beta$; then ${\rm e}^{i(\gamma-\beta)}$ must be real. This implies
that we can choose:
\be
m=\rho, \,\,\,\,\ {\widetilde m}=\epsilon \rho ,\,\,\
\epsilon={\pm 1}, \,\,\,\,\ F_0=f_0,
\label{fivev}
\ee
without loss of generality. Substituting (\ref{fivev}) in
(\ref{fivei}) leads to:
\be
{1 \over b_{11}}\rho \Bigl( \rho^2+b_{11}|a|^2+{ b_{01}\epsilon f_0
\over {\sqrt 2}}\Bigr)=0,
\label{fivevi}
\ee
with two possibilities:
\bea
& & {\rm i}) \,\,\ \rho=0,\\
\label{fivevii}
& & \nonumber \\
& & {\rm ii}) \,\,\ \rho^2=-b_{11}|a|^2+{ b_{01}\epsilon f_0 \over {\sqrt
2}}>0.
\label{fiveviii}
\eea
To determine whether (5.7) or (5.8) is favored we need
 to compute the full potential. Note however that
$b_{11}={1 \over 4\pi} {\rm Im}\,\ \tau_{11}$
is always positive, and therefore (5.8) determines
 a region in the $u$-plane where the monopoles
acquire a vacuum expectation value (VEV).
Depending on the sign of $b_{01}$ we choose the sign of
$\epsilon$. In fact we can replace (5.8)
by:
\be
\rho^2=-b_{11}|a|^2+{1\over {\sqrt 2}}|b_{01}|f_0>0
\label{fiveix}
\ee
and $f_0$ is always measured in units of $\Lambda$.
Thus for the numerical plots we set $\Lambda =1$.
Inserting (5.8) into (\ref{threexxv}) we obtain:
\be
V=-{2 \over b_{11}} \rho^4-{{\rm det}b \over b_{11}}f_0^2
\label{fivex}
\ee
This is good news. It implies that the
 region where the monopoles acquire a VEV
 is energetically favored, and we have confinement.
 Depending on the sign of $b_{01}$,  $m$ and
$\widetilde m$ are either aligned or antialigned.
 The $SU(2)_R$ symmetry of $N=2$ supersymmetry
is broken by the explicit off-diagonal term
$b_{01}m {\widetilde m}/b_{11}$ in (\ref{threexxv})
and by the VEV $\rho \not= 0$.

Where $\rho^2 \rightarrow 0$, the potential maps smoothly onto the potential
for the Higgs region,
\be
V^{(h)}=-{{\rm det}b^{(h)} \over b^{(h)}_{11}}f_0^2,
\label{fivexia}
\ee
since, we recall, ${{\rm det}b/b_{11}}$
is monodromy-invariant. In the monopole region, a
nonzero monopole VEV is favoured, and the effective
potential is given by (\ref{fivex}) and written in
terms of magnetic variables:
\be
V^{(m)}=-{2 \over b^{(m)}_{11}} \rho^4-{{\rm det}b^{(m)}
 \over b^{(m)}_{11}}f_0^2
\label{fivexib}
\ee
where $b^{(h)}$, $b^{(m)}$ are given in
(\ref{threex}), (\ref{threexi}), (\ref{threexxvi}).

 In the Higgs region, the
 effective potential is given by (\ref{fivexia}) and
\begin{figure}
\centerline{
\hbox{\epsfxsize=6cm\epsfbox{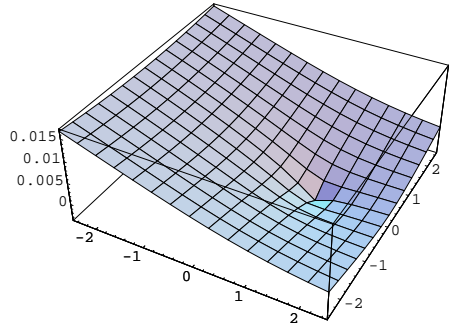}} }
\caption[]{Effective potential, $V^{(h)}$, (\ref{fivexia}).}
\figlabel\figone
\end{figure}
 \begin{figure}
\centerline{
\hbox{\epsfxsize=6cm\epsfbox{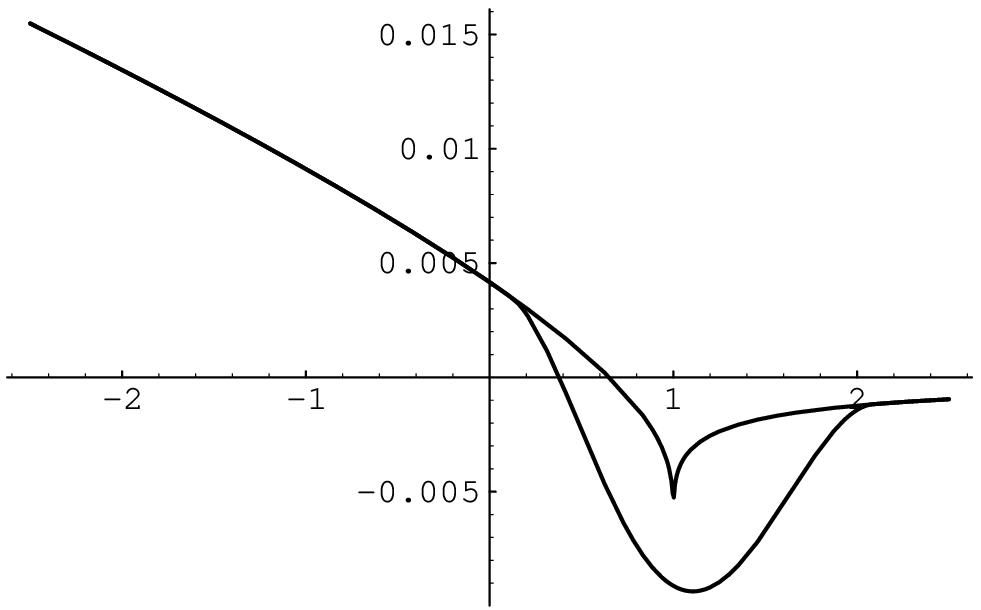}}\qquad
\hbox{\epsfxsize=6cm\epsfbox{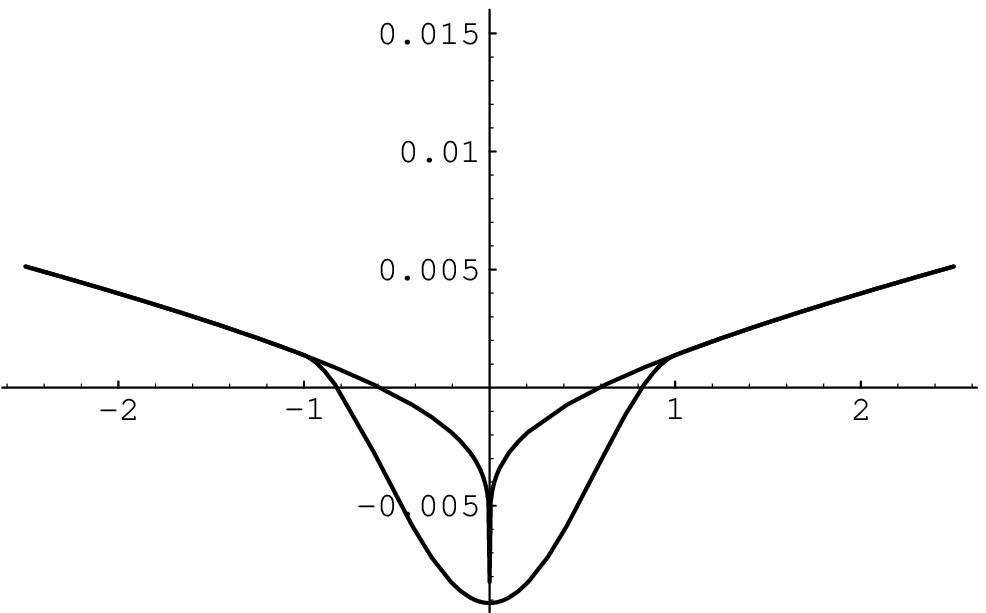}} }
\caption[]{Effective potential, $V^{(h)}$, (\ref{fivexia}) (top) and,
$V^{(m)}$, (\ref{fivexib}) (bottom) along
the real axis (left) and for $u=\Lambda^2(1+i y)$ (right). Both are plotted for
$f_0=0.3\Lambda$.}
\figlabel\figonea
\end{figure}
we plot it in \fig\figone. It has no minimum outside the monopole region near
$u=\Lambda^2$ (where, as we shall see, the energy can be further lowered by
giving the monopoles a VEV). One sees that the shape of
the potential makes the fields roll
towards the monopole region.
In \fig{\figonea}, we plot slices of the potential $V^{(h)}$ along the real
$u$-axis and parallel to the imaginary $u$-axis with ${\rm Re}(u)=\Lambda^2$.
For comparison, we also plot $V^{(m)}$. Note that they agree in the Higgs
region (where the monopole VEV vanishes), and that $V^{(m)}$ lowers the energy
(and smooths out the cusp in $V^{(h)}$ at $u=\Lambda^2$) in the monopole
region.

Next we look at the monopole
region (\ref{fiveix}). $a$ ({\it i.e.} $a^{(m)}$) is a good
 coordinate in this region vanishing at $u=\Lambda^2$.
 As soon as $f_0$ is turned on monopole condensation
 \figalign{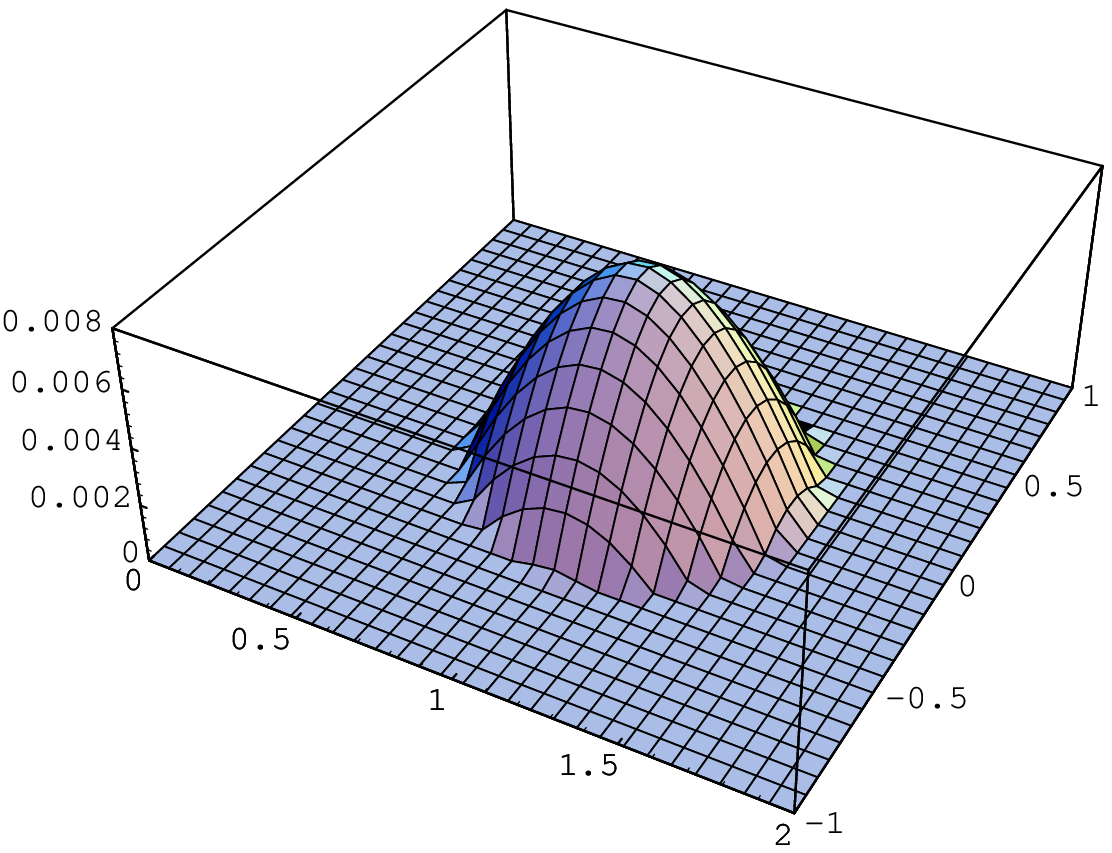}{\figtwo}{Monopole expectation value $\rho^2$ for
$f_0=0.1\Lambda$ on the $u$-plane.}{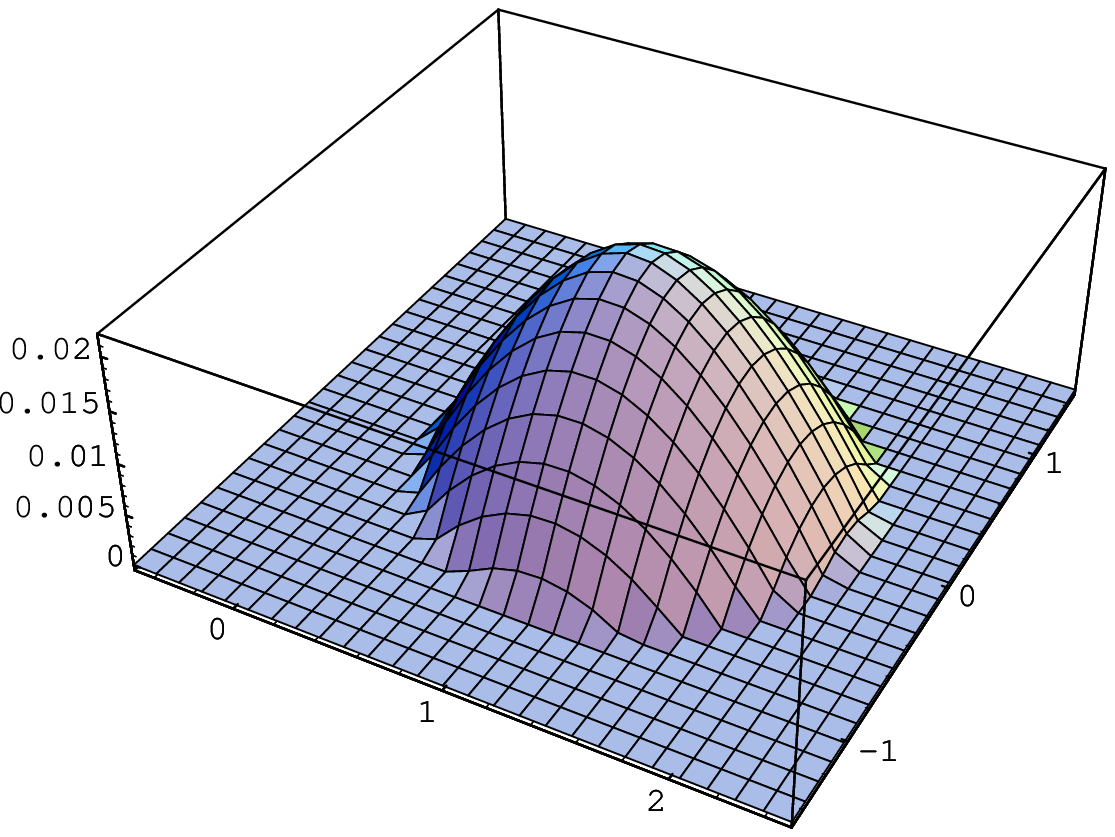}{\figthree}{Monopole
expectation value $\rho^2$ for $f_0=0.3\Lambda$ on the $u$-plane.}
and confinement occur. In \figs{\figtwo,\figthree}\ we
plot $\rho^2$ in the $u$-plane
for values of $f_0 = 0.1 \Lambda$,
$0.3 \Lambda$;
\figalign{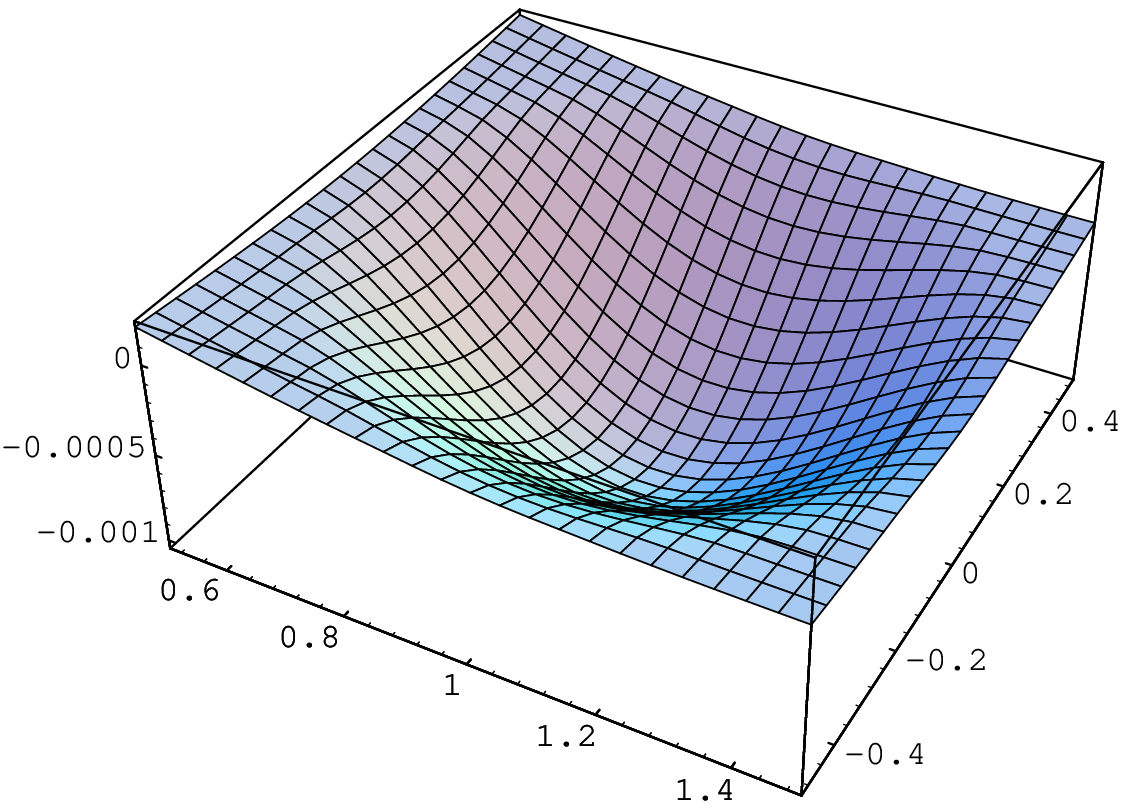}{\figfour}{Effective potential (\ref{fivexib}) for
$f_0=0.1
\Lambda$.}{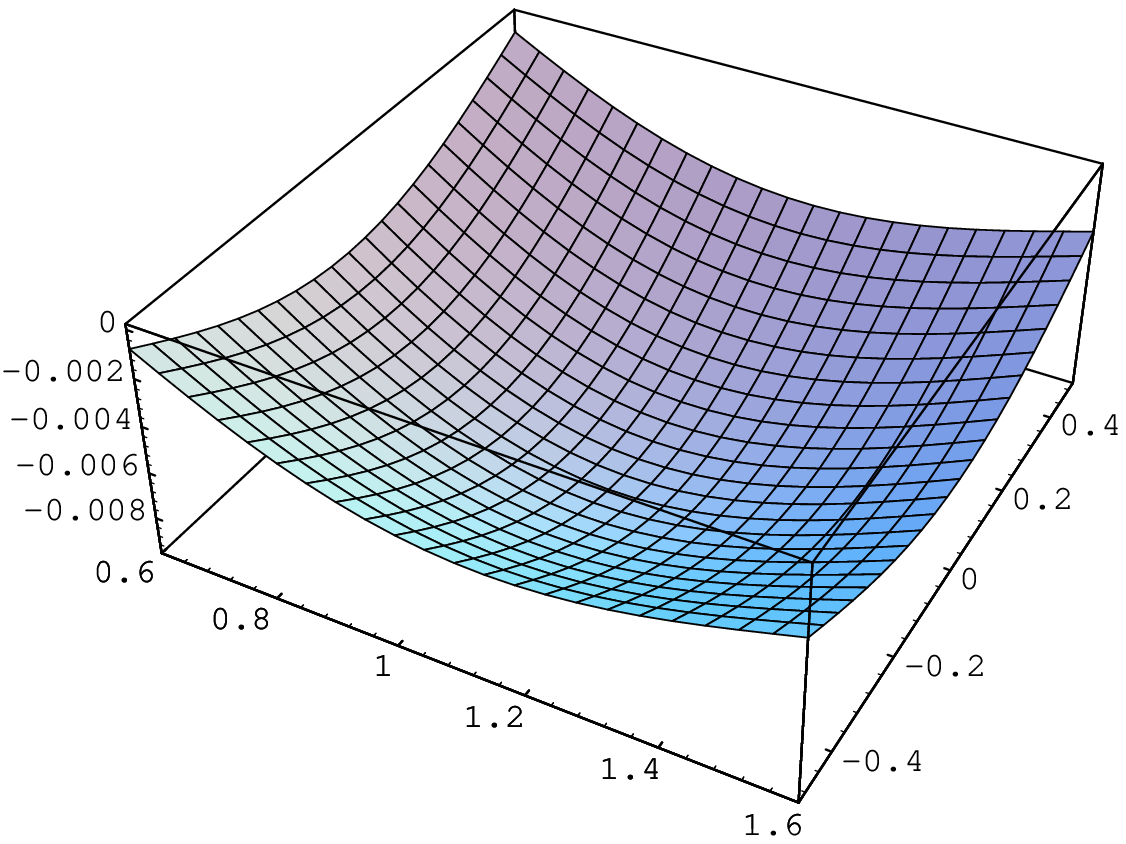}{\figfive}{Effective potential (\ref{fivexib}) for
$f_0=0.3 \Lambda$.}
and in \figs{\figfour,\figfive}\
the effective potential (\ref{fivex}) for
the same values of the supersymmetry
 breaking parameter $f_0$.

One can see that the
minimum is stable and
that the size of the monopole VEV is $\sim f_0$.
There are two features
worth noticing. The first is that the absolute
minimum occurs along
the real $u$-axis. This is seen numerically
and also as
a consequence of the reality properties of
the elliptic functions. Second, as $f_0$ is increased, the region where
(\ref{fiveix}) holds becomes wider.
\figalign{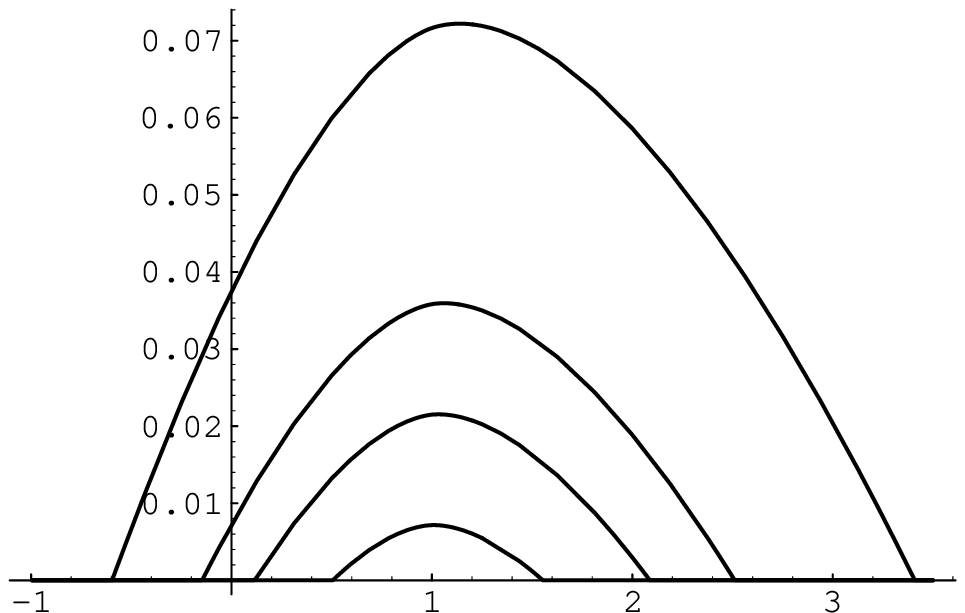}{\figsix}{Plot of $\rho^2$ along the real $u$-axis,
for $f_0/\Lambda=$ (from bottom to top) $0.1$, $0.3$, $0.5$, $1.0$.
}{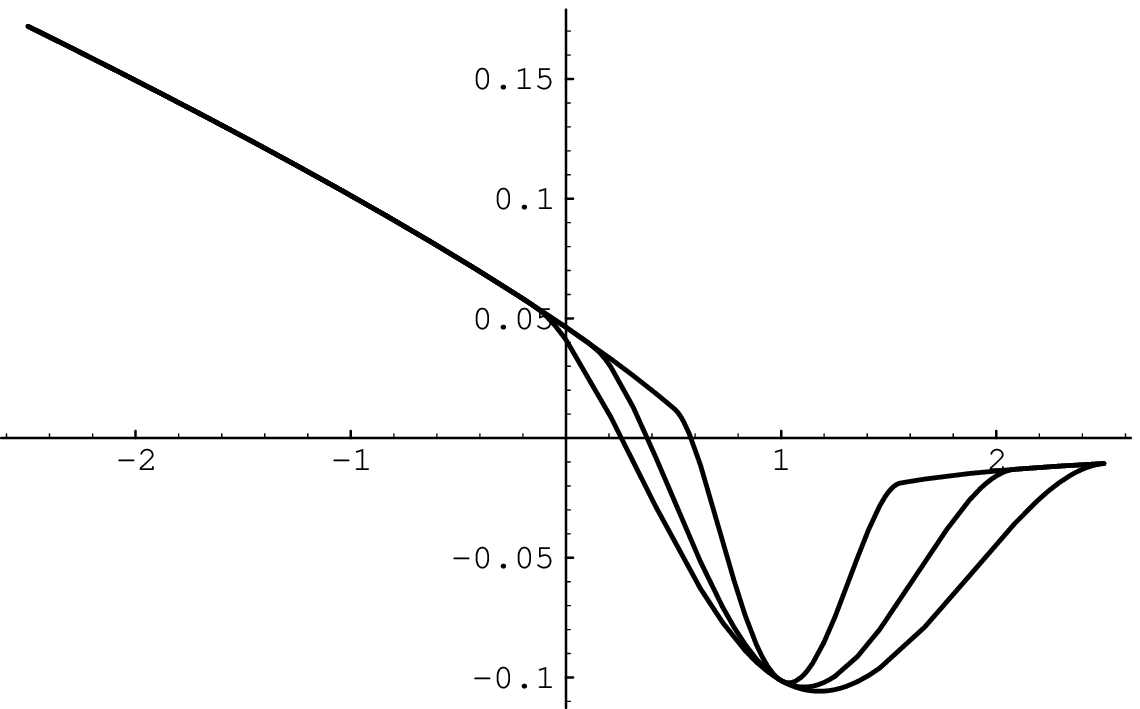}{\figseven}{$V^{(m)}/f_0^2$ along the real $u$-axis for
$f_0=0.1\Lambda$ (top), $0.5\Lambda$ (middle) and $\Lambda$ (bottom).}
 This is seen in \fig{\figsix}, where
$\rho^2$ is plotted along the real $u$-axis as a function of $f_0$.
Accordingly, the minimum of the effective potential moves to the right along
the real
$u$-axis, as one can see in \fig{\figseven}, where
$V^{(m)}/f_0^2$ is plotted for three increasing values of $f_0$ (we have
divided by $f_0^2$ to fit the three potentials on the same graph).

Finally, we turn to the dyon region. To understand what happens in the dyon
region, we study the transformation rules of the $\tau_{ij}$ couplings
under the residual ${\bf Z}_8 \subset U(1)_R$ symmetry whose generator
acts on the $u$-plane as $u \mapsto -u$.  The reason why we need
to analyze in general the behavior under ${\bf Z}_8$ is because
the representation we have chosen for the Seiberg-Witten
solution in sections 2,3 is  well adapted to study the
monopole region. Naively applying them to the dyon region, we may
encounter some discontinuities due to the position
of the cuts. Outside the curve of marginal stability
one can write the prepotential as \cite{swone}:
\be
{\cal F}={i\over 2\pi} a^2 \log{a^2\over \Lambda^2}+
a^2 \sum_{k\ge 1} c_k \Big( {\Lambda \over a}\Big)^{4k}.
\label{prepot}
\ee
If $\omega=e^{2\pi i/8}$ is the generator of the ${\bf Z}_8$
symmetry, it is easy to show that the couplings $\tau_{ij}$
transform according to\footnote{There is one more aspect of the ${\bf Z}_8$
transformation
rules worth noticing. If we implement these rules
we find that the condensate moves to the dyon region,
and one might be tempted to conclude that with this
choice it is the dyon that condenses.  This is not
the case. Using the one-loop $\beta$-function, we know
that $\Lambda^4 \sim {\rm exp}(-{8\pi^2\over g^2}+i\theta)$.
The action of ${\bf Z}_8$ amounts to the change
$\Lambda \mapsto i\Lambda$ or what is the same,
$\theta \mapsto \theta + 2\pi$.  Using the relation
found in \cite{dyon}, when we make this change the massless
state at $u=-\Lambda^2$ (before supersymmetry breaking)
has zero electric charge, while the state at $u=\Lambda^2$
acquires charge one. Thus we find again a monopole
condensate, in a way consistent with the ${\bf Z}_2$-symmetry.
}:
$$
 a\mapsto i a, \qquad a_D \mapsto i (a_D-a),
$$
\be
\tau_{11} \mapsto \tau_{11}-1, \qquad \tau_{01}\mapsto i \tau_{01}, \qquad
\tau_{00}\mapsto -\tau_{00}.
\label{symmetry}
\ee
So the relation between the dyon and monopole variables is:
\bea
a^{(d)}(u)=i a^{(m)}(-u),&\quad
a_D^{(d)}(u)=i \left(a_D^{(m)}(-u)-a^{(m)}(-u)\right),\label{dymon}\\
\tau_{11}^{(d)}(u)=\tau_{11}^{(m)}(-u)-1,&\quad\tau_{01}^{(d)}(u)=i
\tau_{01}^{(m)}(-u),\quad\tau_{00}^{(d)}(u)=-\tau_{00}^{(m)}(-u).\nonumber
\eea
Using the expressions for the monopole couplings in (\ref{threexi}), which are
well-behaved near $u=\Lambda^2$, we obtain expressions for the dyon couplings
which are well-behaved near $u=-\Lambda^2$.
The analysis of (\ref{fiveix}) changes
crucially once these rules are implemented. Near the
monopole region $a^{(m)}\sim i(u-\Lambda^2)$, hence $\tau^{(m)}_{01}\sim i$
is purely imaginary.  In (\ref{fiveix}) although $b_{11}$ diverges
at $u=\Lambda^2$ the divergence is cancelled by the vanishing of
 $a^{(m)}$ at the same point.  Since ${\rm Im}\tau^{(m)}_{01} >0$ as soon
as $f_0\ne 0$ the monopoles condense.  Using (\ref{dymon}), however, we see
that $a^{(d)}\sim (u+\Lambda^2)$ with a real
coefficient.  Thus ${\rm Im}\tau_{01}^{(d)} =0$ at $u=-\Lambda^2$
and we conclude from (\ref{fiveix}) that the dyon condensate {\it vanishes}
along the real $u$-axis. Nevertheless, a dyon condensate {\it is}
energetically favoured in a pair of complex-conjugate regions in the $u$-plane
centered about $u=-\Lambda^2$.
\figalign{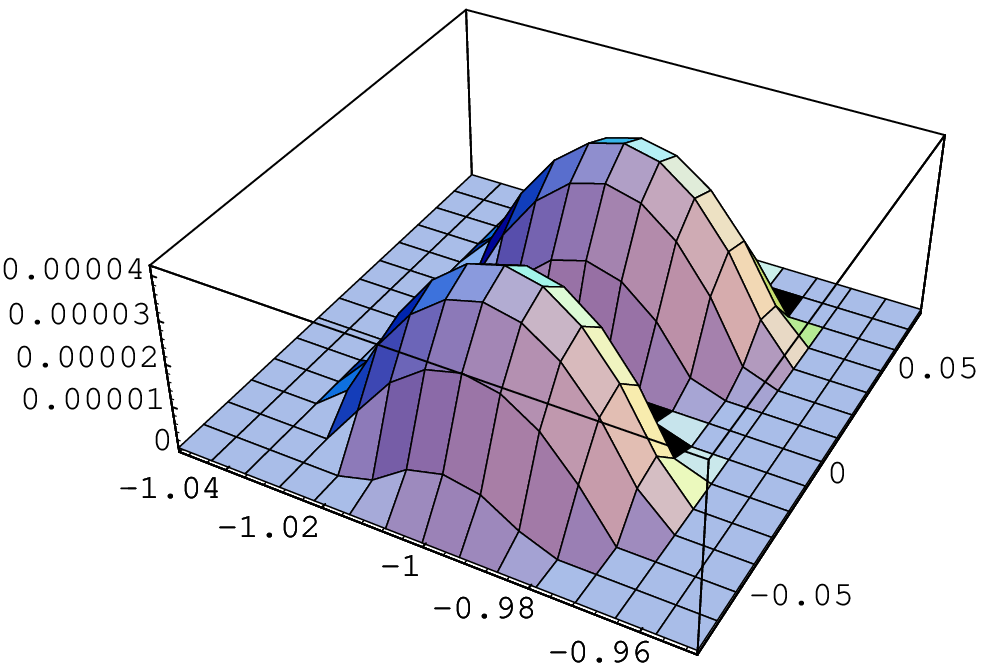}{\figeight}{Dyon expectation value $\rho_{(d)}^2$ for
$f_0=0.3\Lambda$ on the $u$-plane.}{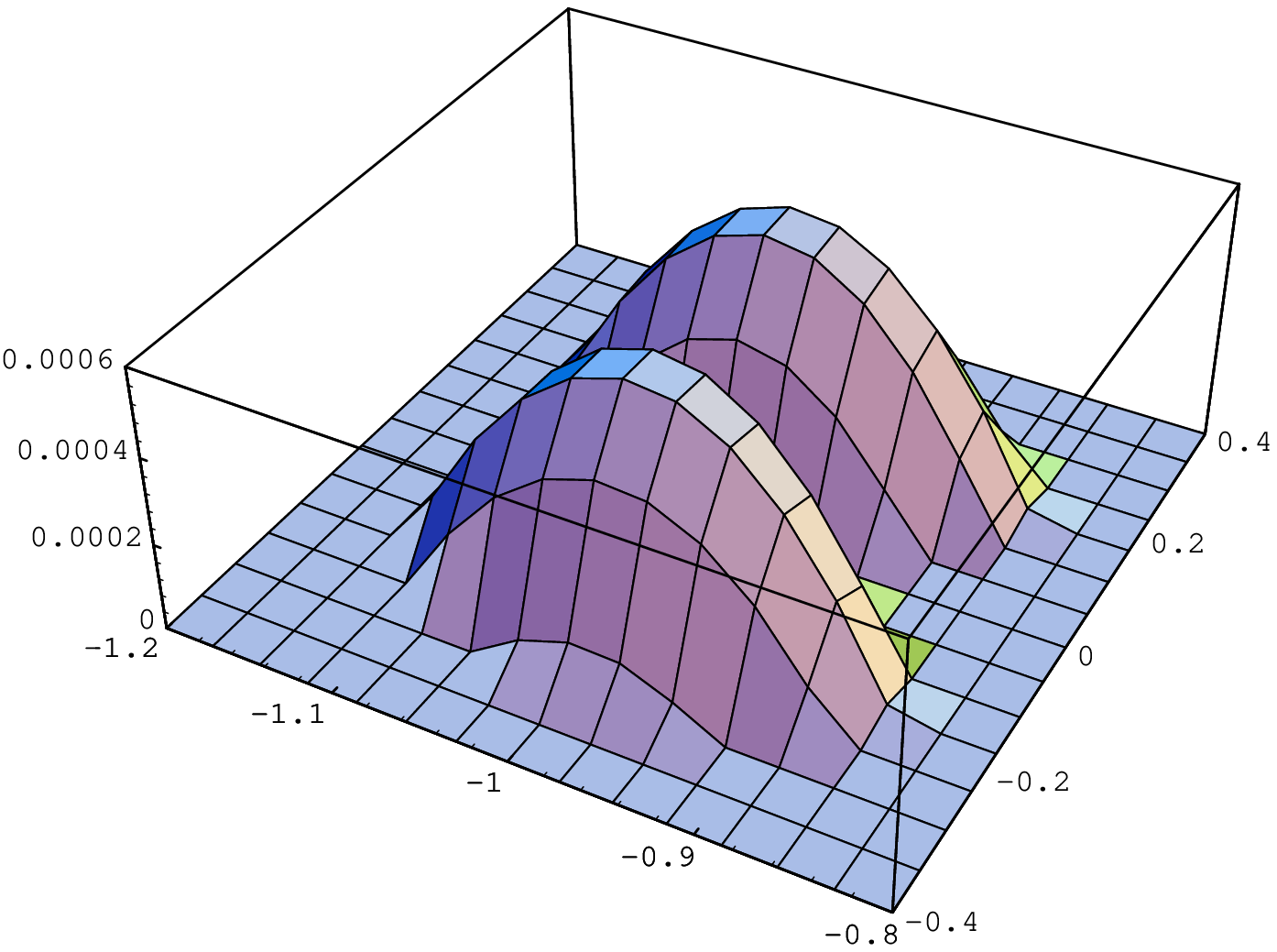}{\figeightb}{Dyon
expectation value $\rho_{(d)}^2$ for $f_0=\Lambda$ on the $u$-plane.}
We plot $\rho_{(d)}^2$, for two different values of $f_0$ in
\figs{\figeight,\figeightb}.

Unlike the monopole VEV, the magnitude of the dyon VEV is {\it tiny} on the
scale of $V^{(h)}$. It therefore makes an all-but-negligible contribution to
the
\begin{figure}
\epsfxsize=6cm
\centerline{\epsfbox{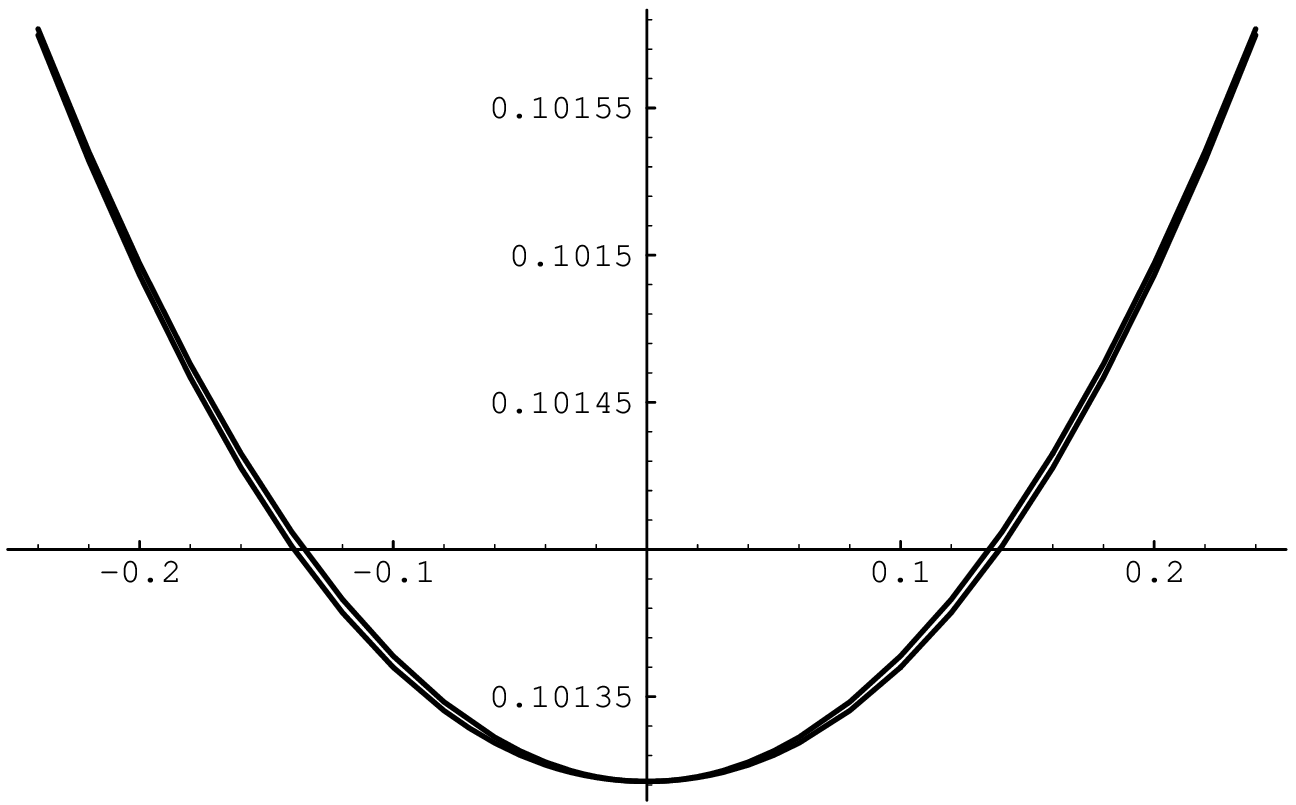} }
\caption[]{Plot of $V^{(h)}(u)$ (top) and $V^{(d)}(u)$ (bottom) versus ${\rm
Im}(u)$ for ${\rm Re}(u)=-\Lambda^2$ and $f_0=\Lambda$.}
\figlabel\figeightc
\end{figure}
effective potential (\fig\figeightc). In particular, $V^{(d)}$ does {\it not}
have a minimum in the dyon region. The only minimum of the full effective
potential is the one we previously found in the monopole region.

Given that the expectation value of the dyons are about two
orders of magnitude smaller than the monopole expectation
value, one might worry that small corrections to the potential
may erase the dyon VEV altogether.  In particular we can
consider the two extra soft breaking terms appearing in
(4.24) in the decoupling of supergravity. Identifying
$m_{3/2}$ with $f_0$, and taking into account that
$i\Sigma\approx 10^{2}$, it is not difficult to include
these effects in our equations for the VEV's or monopoles
and dyons. What we find is that the effect is rather small
and that the expectation values remain essentially the same.
This means that within our approximations, the two types
of VEV do not change significantly once these extra soft
breaking terms are included.

As we have already noted, the monopole region (in which
$\rho_{(m)}^2\neq0$) expands as $f_0$ is increased. Eventually, for $f_0\sim
1.3\Lambda$, it reaches the dyon region (in which
$\rho_{(d)}^2\neq0$). At this point, it is clear that our whole approximation
of including {\it just} the monopole field (or {\it just} the dyon field) in
the effective action breaks down.

What are the other limitations of our
approximations? First, we have neglected  certain soft supersymmetry breaking
terms which  arise in the supergravity action. As discussed in
section 4, these scale to zero in the rigid limit, that is, they are
suppressed
by powers of ${\rm log}{\Lambda \over M_{\rm Pl}}$ or
$\Lambda \over M_{\rm Pl}$ and, for our purposes are negligible. We have also
neglected higher-spinor-derivative corrections to the Seiberg-Witten
effective action. These clearly cannot affect the vacuum structure
in the supersymmetric limit. They also,
{\it by definition} must be supersymmetric; otherwise they lead
to explicitly
hard supersymmetry breaking terms, which is
an entirely different matter from the soft
supersymmetry breaking we are considering. Nevertheless, once supersymmetry
is broken, they can, in principle, lead to corrections to the scalar
potential suppressed by higher powers of $f_0^2/\Lambda^2$. For the moderate
values of $f_0$ that we are considering, these corrections are numerically
rather small, and do not affect the qualitative features of the solutions
 we have found. {\it A priori}, if the higher spinor derivative terms in the
Seiberg-Witten effective action were known, we could systematically improve
our approximations by going to higher order in $f_0^2/\Lambda^2$.

However, the fundamental obstacle to pushing our approximation to larger values
of the soft supersymmetry breaking parameters would remain. The mutual
non-locality of the monopoles and dyons leads to our inability to
calculate the effective potential where the monopole and dyon regions overlap.
Since this is, at least initially, far from the monopole vacuum, we
expect that the monopole vacuum persists, at least as metastable minimum, even
beyond the critical value of $f_0$. But we do not know when (or if) a new,
lower minimum develops once the monopole and dyon regions overlap. If a new
vacuum does appear there, then we would have a first order phase transition to
this new confining phase\footnote{In theories with matter, as discussed in
section 6, this phase transition would change the exotic pattern of chiral
symmetry realized in the monopole and dyon vacua into the standard pattern
expected in the true QCD vacuum.}. This raises the  exciting possibility that
the correct description of the QCD vacuum requires the introduction of
mutually non-local monopoles and dyons. Phases of this nature have been shown
to arise in the
$N=2$
 moduli space for gauge group
$SU(3)$ (see the paper by Argyres and Douglas in \cite{groups}).
Perhaps the way to approach the true QCD vacuum in the
correct phase is to start with one of these $N=2$-superconformal field theories
and turn on a relevant, soft supersymmetry-breaking perturbation.

Although we have illustrated our method of supersymmetry breaking so
far
only for pure $SU(2)$, the fact that the soft breaking terms are all
produced by making $\Lambda$ a function
of the spurion makes this procedure
quite universal, and similar results can be obtained for other gauge
groups with and without quark hypermultiplets with arbitrary masses.
One example is illustrated in the next
section where we include two doublets of
massless quarks.

\section{Including Two Massless Quark Multiplets}
\setcounter{equation}{0}

When $N_f$ massless hypermultiplets of quarks are included the global
flavour
symmetry is $O(2N_f)$, because the ${\bf 2}$ and the ${\bf {\bar 2}}$
representations of $SU(2)$ are equivalent. The full group of global
symmetries
is $O(2N_f) \times SU(2)_R \times U(1)_R$. In \cite{swtwo} Seiberg
and Witten
have given the exact form of the low-energy effective action when
$N_f \le 4$
with and without masses. When $N_f=2$ and the masses are set to zero,
 the global symmetry is $O(4) \times SU(2)_R \times U(1)_R$, and the
elliptic curve
is exactly (\ref{xiii}):
\be
\label{sixi}
y^2=(x^2-\Lambda^4)(x-u).
\ee
The reason is that the normalizations of \cite{swone}
and \cite{swtwo} are different. In
\cite{swone} the charge operator is normalized so that the
$W^{\pm}$-boson
has charge $\pm 1$, while in \cite{swtwo} the quarks are taken to
have
charges $\pm 1$ and hence $W^{\pm}$ has charge $\pm 2$. In the
conventions of \cite{swtwo} the curve associated to the $N_f=0$ case
is:
\be
y^2=x^2(x-u)+{1 \over 4}\Lambda^4x,
\label{sixii}
\ee
and the monodromy group is contained in $\Gamma_0(4)$. Using the
curve (\ref{sixi})
with the conventions of \cite{swtwo} most of the formul\ae\ of sections
2, 3 still apply. There are two singularities in the moduli space, at
$u={\pm}\Lambda^2$. However now the monopoles and dyons behave
 respectively
as $({\bf 2},{\bf 1})$ and $({\bf 1},{\bf 2})$ with respect to the
global
$O(4)$ group. In the monopole region the full global symmetry group
is
 $SU(2)_r \times SU(2)_R \times U(1)_R$ with $SU(2)_r \subset O(4)$,
and
similarly in the dyon region changing $SU(2)_r \longleftrightarrow
SU(2)_l$.
We can arrange the scalar monopoles in a $2 \times 2$ matrix:
\be
\Phi=\left(\begin{array}{cc} m_1&{\widetilde m}_1^{*}\\
                           m_2&{\widetilde m}_2^{*}\end{array}\right),
\label{sixiii}
\ee
which under $SU(2)_r \times SU(2)_R$ transforms according to:
\be
\Phi \rightarrow g_r \Phi g_R^{-1}.
\label{sixiv}
\ee
Making $\Lambda$ a function of the spurion $S$, and again for
simplicity setting $D_0=0$, we obtain a potential analogous
to (\ref{threexxv}):
\bea
V &=& {1 \over 2b_{11}} \big( |m|^2-|\widetilde m|^2 \big)^2 + 2|a|^2
 \big( |m|^2+|\widetilde m|^2 \big) \nonumber \\
 & &\nonumber\\
&+& {1 \over b_{11}}\Bigl| b_{01}f_0+ {\sqrt 2} m \cdot {\widetilde m}
\Bigr|^2
-b_{00}f_0^2,
\label{sixv}
\eea
where $|m|^2=|m_1|^2+|m_2|^2$, $ m \cdot {\widetilde m}=m_1{\widetilde
m}_1+m_2{\widetilde m}_2$, and phases have been chosen to make $f_0$ real.
If we use the identity:
\be
\sum_{i} \big( {\rm Tr}\ \sigma_i A \big)^2=2{\rm Tr}\
A^2-\big({\rm Tr}\ A \big)^2,
\label{sixvi}
\ee
which holds for any $2\times 2$ matrix of the form $A=a_0+a^i
\sigma_i$, with $\sigma_i$ the Pauli matrices, the potential $V$ can
be written in a more
transparent form:
\bea
V &=& {1 \over 2b_{11}} \Big(2{\rm Tr}( \Phi^{\dagger}\Phi)^2
-({\rm Tr}\
\Phi^{\dagger}\Phi )^2  \Big) +
2|a|^2{\rm Tr}\ \Phi^{\dagger}\Phi   \nonumber \\
& &\nonumber\\
&+& {{\sqrt 2}b_{01}\over b_{11}}  f_0 {\rm Tr}\ \sigma_1
\Phi^{\dagger}\Phi -{{\rm det}\ b \over b_{11}}f_0^2.
\label{sixvii}
\eea
Varying $V$ with respect to $\Phi^{\dagger}$ leads to:
\be
{1 \over b_{11}} \Big(2 \Phi \Phi^{\dagger} \Phi -({\rm Tr}\
\Phi^{\dagger}\Phi )\Phi  \Big) +
2|a|^2 \Phi + {{\sqrt 2}  b_{01} \over b_{11}}f_0  \Phi \sigma_1 =0.
\label{sixviii}
\ee
Multiply (\ref{sixviii}) by $\Phi^{\dagger}$, and let $A \equiv
\Phi^{\dagger} \Phi$:
\be
 {1 \over b_{11}} \Big(2A^2 -({\rm Tr}\
A )A \Big) +
2|a|^2 A + {{\sqrt 2}  b_{01} \over b_{11}}f_0 A \sigma_1 =0.
\label{sixix}
\ee
There are several solutions to (\ref{sixix}).

i) If $m_i=0$, $\Phi^{\dagger} \Phi=\left(\begin{array}{cc} 0&0\\
                          0&|\widetilde m|^2\end{array}\right)$, and
(\ref{sixix})
implies ${\widetilde m}_i=0$. The same conclusion applies if ${\widetilde
m}_i=0$. Hence $\Phi=0$ and the monopoles do not get a VEV.
 As in section 5, this phase has higher energy.

ii) If the matrix $\Phi$ is invertible, so is $A$. We can
left-multiply
by $A^{-1}$ in (\ref{sixix}), and then take the trace.
 This implies that $a=0$. In the monopole region this means
$u=\Lambda^2$. At this particular point (\ref{sixix}) implies:
\be
|m|^2=|{\widetilde m}|^2, \,\,\,\,\,\,\ m\cdot{\widetilde
m}=-b_{01}f_0/{\sqrt 2}.
\label{sixx}
\ee
In this branch the monopole acquire a VEV, but their auxiliary fields
 do not, while the auxiliary fields of $a$ get a VEV. When
(\ref{sixx}) is
inserted in (\ref{sixvii}) we obtain:
\be
V=-b_{00}f_0^2,
\label{sixxi}\ee
where $b_{00}$ is evaluated at $a=0$. We will comment on this branch
later.

iii) Finally, $\Phi$ may not be invertible. Ignore the cases
$m_i={\widetilde m}_i=0$ covered in i); $m_i$ and ${\widetilde m}_i$ must be
proportional:
\be
m_i={\lambda}^{-1}{\widetilde m}_ i,\,\,\,\,\,\ \lambda \not= 0.
\label{sixxii}
\ee
Thus,
\be
\Phi=\left(\begin{array}{cc} m_1&\lambda m_1 \\
                           m_2&\lambda
m_2\end{array}\right),\,\,\,\,\,\,\ \Phi^{\dagger} \Phi= |m|^2
\left(\begin{array}{cc} 1&\lambda\\

{\lambda}^{*}&|\lambda|^2\end{array}\right).
\label{sixxiii}
\ee
(\ref{sixix}) now implies that $\lambda = \epsilon =\pm 1$; and
\be
{2 \over b_{11}}|m|^4+2|a|^2|m|^2+
{\sqrt 2}{ b_{01}\epsilon f_0 \over b_{11}}|m|^2=0.
\label{sixxiv}
\ee
Note however that we have already encountered (\ref{sixxiv})
in the previous
 section (see (\ref{fivevi}), (\ref{fiveix})),
and we will not repeat the analysis here.
Away from $u=\Lambda^2$ we have one ground state.
 The symmetry of (\ref{sixvii}) is not all of $SU(2)_r \times
SU(2)_R$
because the term ${\rm Tr} \sigma_1 \Phi^{\dagger}\Phi$ breaks
explicitly
$SU(2)_R$ to the $U(1)$ subgroup commuting with $\sigma_1$. If we had
also
included a $D_0$ soft breaking term, $SU(2)_R$ would be completely
broken.
However with only $f_0 \not= 0$ the global group
$SU(2)_r \times U(1)_R$ breaks to $U(1)$ because of the VEV for the
monopoles.
With $D_0 \not= 0$ we would have $SU(2)_r$ breaking completely while
$SU(2)_l$
remains intact. If we restrict the
computation to regions where $f_0/\Lambda < 1$ we can use the effective
action to obtain the Goldstone boson effective
lagrangian
up to two derivatives, including the non-perturbative corrections. Once quark
masses are included this may be an
interesting ground to test many ideas about the computation of the
low-energy
chiral lagrangian in terms of QCD.

To obtain the standard pattern
of chiral symmetry breaking, in which $SU(2)_l\times SU(2)_r\to SU(2)_V$, we
presumably need to be in the phase, alluded to in the previous section, where
both monopoles and dyons condense.

The phase ii) is analogous to the the two Higgs phases in the $N_f=2$
case
described in \cite{swtwo}. In the purely
 supersymmetric setting at the classical level, there are together
with the
Coulomb phase two Higgs phases meeting at the origin of the classical
moduli
space. In the quantum theory these two phases meet the Coulomb phase
 at different points. This is precisely what is found in solution
ii): there
are two analogues of the Higgs phase attached to either $u=\Lambda^2$
or
$u=-\Lambda^2$. These two Higgs branches lie on a flat direction of
the effective
potential, where $V$ takes the constant value given by (\ref{sixxi}).
\begin{figure}
\epsfxsize=6cm
\centerline{\epsfbox{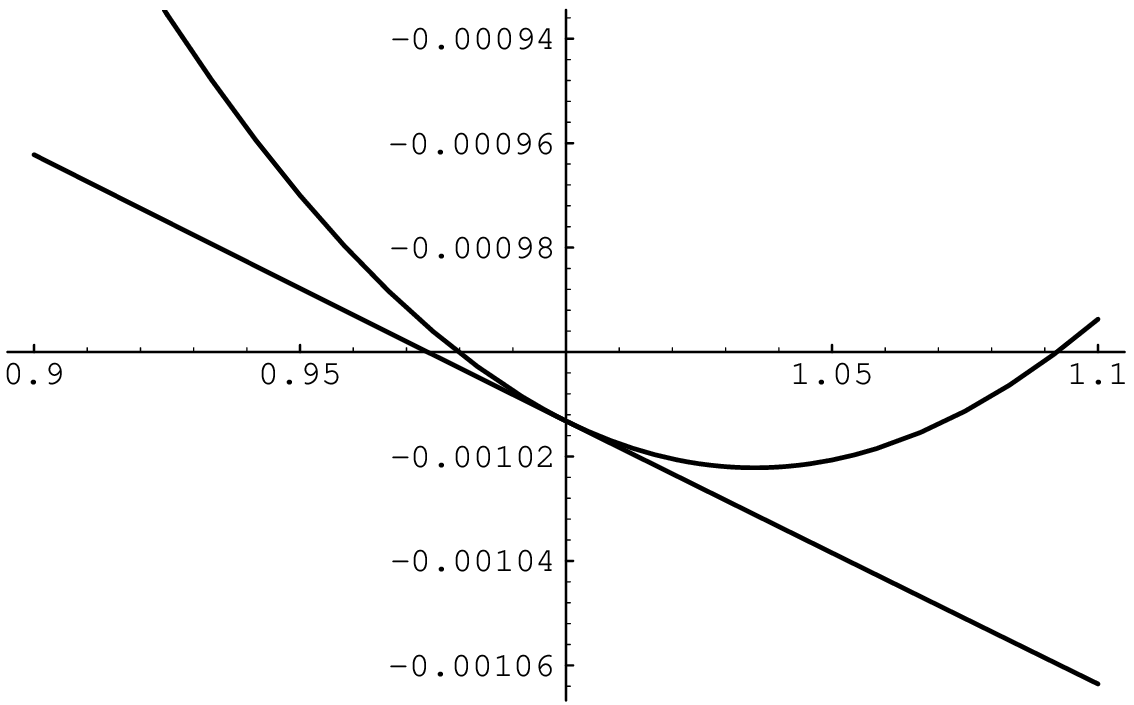} }
\caption[]{Plot of (\ref{fivexib}) (top) and (\ref{sixxi}) (bottom) near
$u=\Lambda^2$
for $f_0=0.1\Lambda$.}
\figlabel{\fignine}
\end{figure}
Notice that (\ref{fivexib}), when evaluated in $a=0$, equals
(\ref{sixxi}) (as one can see in \fig{\fignine}\ for $f_0=0.1\Lambda$).
Hence there are no discontinuities in the vacuum energy
and both phases are smoothly connected, as one should expect
in a theory with matter fields in the fundamental representation \cite{higgs}.
As the minimum of (\ref{fivexib}) lies on ${\rm Re}u >1$,
${\rm Im}u=0$ for any non-zero $f_0$, the phase in iii) is
 energetically favoured.

\section{Conclusions}

In this paper we have shown that there is a general procedure to
softly
break $N=2$ down to $N=0$ theories without losing the holomorphic
properties
of the Seiberg-Witten solution \cite{swone,swtwo}. When the
supersymmetry breaking scale is small compared to the dynamical scale
 $\Lambda$, this leads to an analytic determination of the low-energy
effective
action including non-perturbative effects.
The advantage of breaking softly using a dilaton spurion is its
universality:
 it applies to any of the generalizations of
\cite{swone,swtwo} in \cite{groups}, and in particular to
 theories with massive quarks.

We have exhibited two applications to
$N=2$ theories with gauge group $SU(2)$ and $N_f=0,2$, exhibiting
some details
of their phase structure and patterns of symmetry breaking. We have
also shown
that the structure of the soft-breaking terms induced can be derived
from a
 spontaneously broken $N=2$ supergravity theory.
One could envisage more complicated ways of achieving similar
results.
The basic idea is to have an extra $N=2$ vector multiplet
 invariant under the Seiberg-Witten monodromy.
Thus we could consider embedding the $SU(2)$ moduli space into the
$SU(3)$
moduli space, and determine the $SU(3)$ vector multiplet in the
low-energy theory with this property; and then declare this
multiplet to become the spurion. While feasible, this is
not straightforward due to the subtleties in embedding
the Seiberg-Witten moduli space inside the $SU(3)$ or higher moduli
spaces.

It is intriguing that the clear breakdown of our approach is associated with
the coalescence of the two regions in which, respectively, the monopoles and
dyons condense. Though monopole condensation is clearly the
mechanism of confinement for small values of the soft supersymmetry-breaking
parameters, it appears likely that nature of the QCD vacuum in the decoupling
limit is more complicated, involving, perhaps, the condensation of both
monopoles and dyons.

There are many issues in quantum field theory which we believe can be
explored
with this method. In particular one can obtain the dependence in
quark masses
in the low-energy Goldstone boson Lagrangian (for the time being with
a
non-QCD-like pattern of symmetry breaking), and one can analyze the
non-perturbative
ambiguities appearing in the Operator
Product Expansion associated to renormalon problems.
 It would also be interesting to study the large-$N$ limit. In $N=2$
Yang-Mills
theories the large-$N$ limit is very rich and by including $N$ in our
scaling
relations it may be possible to reach reliably more realistic
scenarios. We
plan to return to these issues in the future.

Some years ago it was almost inconceivable to expect analytic control
on
fully interacting four-dimensional gauge theories. After Seiberg and
Witten's
big leap, we hope this work is a small step towards the real world.

\newpage

{\large\bf Acknowledgements}

We have benefitted from discussions with many colleagues. We
would like to thank: E.~\'Alvarez, A.~Cohen, R.~D'Auria, B.~de Wit,
S.~Ferrara, P.~Fr\'e, L. Girardello, C.~G\'omez, S.D.H.~Hsu, R.~Jackiw,
J.I.~Latorre and W.~Lerche. Two of us (L.~A.-G. and J.D.) would like
to
thank the Benasque Center for Physics (Pyrenees) for its hospitality
while this work was begun. We also want to thank R.~D'Auria, P.~Fr\'e
and B.~De Wit for kindly making available to us part of their work
prior to
publication. M.M. would like to thank the Theory Division at CERN
for its hospitality. The work of J.D.~is supported by NSF grant PHY-9511632,
the Robert A.~Welch Foundation and an
Alfred P.~Sloan Foundation Fellowship. The work of C.~K.
is supported by EEC contracts SC1$^{*}$-0394C and SC1$^{*}$-CT92-0789.
The work of M.M.~is supported in part by DGICYT under grant
PB93-0344 and by
CICYT under grant AEN94-0928.

\end{document}